\documentclass[journal]{IEEEtran}
\ifCLASSINFOpdf
\else
\fi
\usepackage{amsmath}
\usepackage{amssymb}
\usepackage[pdftex]{graphicx}
\usepackage{algorithm}
\usepackage{algorithmicx}

\usepackage{algpseudocode}
\usepackage{color}
\usepackage{multirow}
\usepackage{makecell}

\usepackage{cite}
\usepackage[hidelinks]{hyperref}

\usepackage{subfigure}
\usepackage{threeparttable}
\usepackage{stfloats}

\newtheorem{remark}{Remark}
\newtheorem{definition}{Definition}
\begin{document}
%
\title{Reducing Action Space: Reference-Model-Assisted Deep Reinforcement Learning for Inverter-based Volt-Var Control}
%
%

\author{Qiong~Liu,
        Ye~Guo,
        Lirong~Deng,
        Haotian~Liu,
        Dongyu~Li,
        and Hongbin~Sun
\thanks{This work was supported by the National Key R\&D Program of China (2020YFB0906000, 2020YFB0906005).}
\thanks{Qiong Liu, Ye Guo are with the Tsinghua-Berkeley Shenzhen Institute, Tsinghua University, Shenzhen, 518071, Guangdong, China, e-mail: guo-ye@sz.tsinghua.edu.cn.}
\thanks{Lirong Deng is with the Department of Electrical Engineering, Shanghai University of Electric Power, Shanghai, 200000, China}
\thanks{Haotian Liu, Hongbin Sun are with the State Key Laboratory of Power Systems, Department of Electrical Engineering, Tsinghua University, Beijing 100084, China}
\thanks{Donyu Li is with the School of Cyber Science and Technology, Beihang University, Beijing, 100191, China.}
\thanks{Manuscript received XX, 2022; revised XX, 2022.}}

%
%

\markboth{IEEE TRANSACTIONS ON SMART GRID, ~VOL.~XX, NO.~X, January 2022}%
{Shell \MakeLowercase{\textit{et al.}}: Bare Demo of IEEEtran.cls for IEEE Journals}
%



\maketitle

\begin{abstract}
Reference-model-assisted deep reinforcement learning (DRL) for inverter-based Volt-Var Control (IB-VVC) in active distribution networks is proposed.
We \textcolor{red}{investigate} that a large action space increases the learning difficulties of DRL and degrades the optimization performance in the process of generating data and training neural networks.
To reduce the action space of DRL, we design a reference-model-assisted DRL approach.
We introduce definitions of the reference model, reference-model-based optimization, and reference actions.
The reference-model-assisted DRL learns the residual actions between the reference actions and optimal actions, rather than learning the optimal actions directly.
Since the residual actions are considerably smaller than the optimal actions for a reference model, we can design a smaller action space for the reference-model-assisted DRL. 
It reduces the learning difficulties of DRL, and optimises the performance of the reference-model-assisted DRL approach.
It is noteworthy that the reference-model-assisted DRL approach is compatible with any policy gradient DRL algorithms for continuous action problems. 
This work takes the soft actor-critic algorithm as an example and designs a reference-model-assisted soft actor-critic algorithm.
Simulations show that 1) large action space degrades the performance of DRL in the whole training stage, and 2) reference-model-assisted DRL requires fewer iteration times and returns a better optimization performance.
\end{abstract}

\begin{IEEEkeywords}
Volt-Var control, deep reinforcement learning, active distribution network.
\end{IEEEkeywords}


%
\IEEEpeerreviewmaketitle

\section{Introduction}

\IEEEPARstart {T}o achieve a carbon-neutral society, more distributed generations (DGs) will be integrated into active distribution networks (ADNs). 
The high penetration of DGs would cause severe over-voltage problems. Volt-Var control, as an effective method, has been widely integrated into ADNs. 
Recently, most DGs are inverter-based, which can provide reactive power rapidly with a step-less regulation. Inverter-based Volt-Var control (IB-VVC) has attached increasing interest.

Model-based optimization methods are widely used to solve IB-VVC problems. Those methods can solve a reliable solution under the accurate power flow model of ADNs.
However, in a real application, it may be difficult to obtain a high-accuracy model for the distribution system operator. The control performance would decrease with the decrease in model accuracy.

As a model-free method, deep reinforcement learning (DRL) has made breakthrough achievements in computer games, chess go, robots, and self-driven cars, which also attract huge interest in VVC problems\cite{chen2021reinforcement,liu2021bi,yan2020real, sun2021two,Gao_Consensus, DIcao_MADRL_PV, Two_Timescale_yang,haiyueRobust,haotianonline_multi}. 
For VVC problems, DRL has two attractive advantages:
1) It learns to make actions from interaction data, and a precision model is not needed;
2) It has a high computation efficiency that only needs a forward computation of the neural network in the application stage, because the time-consuming optimization process is shifted into the training stage.
However, DRL also suffers from optimality and convergence issues. There are maybe three directions to address the two issues.


Firstly, design a good reward function by trading off the weight of power loss and voltage violation. 
A small weight of voltage violation cannot penalize
the voltages into the normal range, whereas a large weight results in worse or even unstable learning performance \cite{zhang2020deep, Wangwei-Safe-Off-Policy}.
To alleviate the problem, the paper \cite{zhang2020deep} uses a switch reward method to consider the priority of voltage violation. If voltage violations appear, the reward only contains the penalty of voltage violation.
Paper \cite{Wangwei-Safe-Off-Policy} designs a constrained soft actor-critic algorithm to tune the ratio automatically. 
A reward function that is easy to learn improves the DRL performance.
However, there is still a gap between DRL results and the global optimal results.

Secondly, select a suitable DRL algorithm or make specific modifications according to the VVC characteristic. There is maybe no DRL algorithm over others in all tasks, for example, the original paper of soft actor-critic (SAC) only shows SAC outperforms other algorithms in 4 out of 6 tasks \cite{haarnoja2018soft}. Paper \cite{Wangwei-Safe-Off-Policy} also shows that constrained soft actor-critic has better optimality and faster convergence process compared with a constrained proximal policy optimization algorithm. Considering IB-VVC is a single-period optimization problem and has two objectives with different properties, single-period two-critic DRL is proposed 
which results in better optimization and convergence performance \cite{liu2022OSTC}.
In this direction, 
we may need to develop specific and high-performance DRL algorithms by considering more details of VVC properties.
%

Thirdly, utilize a power flow model to assist DRL.
A two-stage deep reinforcement learning framework first trains a robust policy in an inaccurate power flow model, and then the DRL agent converges to large rewards with small iterations when training in a real environment \cite{haotian_Two_Stage}. To improve the data efficiency, model-based DRL learns a power flow model from historical operation data, and then trains the policy on the learned model \cite{gao2022model,Cao_physical}.
To ensure no voltage violation appears in the training process, an optimization algorithm based on the approximated power flow model is designed to readjust the action of DRL when DRL makes unsafe actions \cite{kou2020safe,gao2022model}.
There are many methods to utilize model to improve DRL performance from different perspectives, and it needs continuously research.

As the above discussion, three directions can improve the DRL-based VVC performance: design better reward functions, select or modifies  suitable DRL algorithms, and utilize the power flow model.
This paper follows the direction of utilizing the power flow model and proposes a reference-model-assisted DRL approach. 
The approach reduces the action space by utilizing the solutions of the reference-model-based optimization. It addresses the learning difficulties of DRL due to large action space.
The main contributions of this paper are the following:

\begin{itemize}
\item[1.] We realize that a large action space increases the learning difficulties of DRL. Five reasons are investigated in the process of generating data and training neural networks, which are verified by simulation experiments point-wisely.

\item[2.] We analyze the necessary effective conditions of the reference-model-assisted DRL approach and define the reference model and reference actions from the view of optimization.

\item[3.] 
We design a reference-model-assisted DRL approach to learn the residual optimal actions between the reference actions and the ideal optimal actions. Compared to general DRL results, this approach achieves better VVC performance by selecting a smaller action space.

\item[4.] We discuss the two potential problems of setting the action space of reference-model-assisted DRL, in which the action space is ``too small" or ``too large".
For a ``too small" action space, even though the final action cannot reach the optimal action, reference-model-assisted DRL still can improve optimization performance based on the reference model actions.
For a ``too large" action space, we propose a dynamic mapping technology to readjust the action space to ensure the final actions are always in the action space.
 \end{itemize}

The remainder of the paper is organized as follows. Section II introduces the preliminaries of IB-VVC and problem formulation. Section III investigates the impact of large action space on the learning difficulties of DRL. Reference-model-based optimization for IB-VVC is introduced in section IV. Section V presents the proposed reference-model-assisted DRL approach. Section VI verifies the proposition of Section III and the superiorities of the proposed method of Section V through extensive simulations. Section VI concludes the results.

\section{Preliminaries and Problem Formulation}
This section first introduces the preliminaries of IB-VVC and DRL, and then formulates the IB-VVC problem as a Markov decision process (MDP).

\subsection{Preliminaries of IB-VVC}
IB-VVC minimizes the power loss and eliminates the voltage violation of ADNs by optimizing the outputs of inverter-based devices. 
It is usually formulated as a single-period optimal power flow \cite{dall2015photovoltaic, liu2017distributed}:
\begin{equation}
\begin{split}
   &\min \limits_{\boldsymbol{x},\boldsymbol{u}} r_p(\boldsymbol{x}, \boldsymbol{u}, \boldsymbol{D},\boldsymbol{p},\boldsymbol{A}) \\
s.t. \quad& f(\boldsymbol{x}, \boldsymbol{u}, \boldsymbol{D},\boldsymbol{p},\boldsymbol{A}) =0 \\
& \underline{\boldsymbol{u}} \leq \boldsymbol{u} \leq \bar{\boldsymbol{u}} \\
  \quad & \underline{h}_v \leq h_v((\boldsymbol{x}, \boldsymbol{u}, \boldsymbol{D},\boldsymbol{p},\boldsymbol{A}) \leq \bar{h}_v,
\end{split}
\end{equation}
where $r_p$ is power loss. $\boldsymbol{x}$ is the state variable vector of the ADN like active power injection $\boldsymbol{P}$, reactive power injection $\boldsymbol{Q}$, and voltage magnitude $\boldsymbol{V}$, $\boldsymbol{u}$ is a vector of the control variables which are reactive power produced by static var compensators (SVCs) and IB-ERs, $\boldsymbol{D}$ is the vector of uncontrollable power generations of distributed energy resources and power loads, $\boldsymbol{P}$ is the parameters of  an ADN, $\boldsymbol{A}$ is the incidence matrix of an ADN, $f$ is the power flow equation, $\underline{u}$, $\bar{u}$ are the lower and upper bounds of controllable variable,  and $\underline{h}_v$, $\bar{h}_v$ are the lower and upper bounds of voltage.
This paper consider an ADN with $n+1$ buses, and 
bus $0$ is a subsection connected to an external grid.
The problem is a nonlinear programming (NLP) problem that can be solved by off-the-shelf NLP solvers such as IPOPT solvers \cite{zimmerman2010matpower}. Generally, the power flow equation $f$, the parameter $\boldsymbol{P}$ and the topology information are difficult to be acquired accurately, which may affect the VVC performance.

\subsection{Preliminaries of DRL}

Reinforcement learning (RL) is a data-driven optimization method that learns the policy to maximize the cumulative reward in the environment. We generally model the problem as a Markov decision process (MDP).
At each step, the DRL agent observes a state $\boldsymbol{s}_t$ and generates an action $\boldsymbol{a}_t$ according to the policy $\pi$. After executing the action to the environment, the DRL agent observes a reward $r_t$ and the next state $\boldsymbol{s}_{t+1}$.
The process generates a trajectory $\tau = \left(\boldsymbol{s}_{0}, \boldsymbol{a}_{0}, r_{1}, \boldsymbol{s}_{1}, \boldsymbol{a}_{1}, r_{2}, \ldots\right)$.
The infinite-horizon discounted cumulative reward of RL agent obtained is  $R(\tau)=\sum_{t=0}^{\infty} \gamma^{t} r_{t}$, where $\gamma$ is the discounted factor with $0 \leq \gamma< 1$.
DRL trains the policy $\pi$ to maximize the expected infinite horizon discounted cumulative reward:
\begin{equation}
\pi^{*}=\arg \max _{a \sim \pi} \mathbb{E} [R(\tau)].
\end{equation}

In value-based or actor-critic RL, state-action value function  $Q^{\pi}(\boldsymbol{s},\boldsymbol{a})$ is defined to evaluate the performance of policy. 
$Q^\pi(\boldsymbol{s},\boldsymbol{a})$ is the expected discounted cumulative reward for starting in state $\boldsymbol{s}$, taking an arbitrary action $\boldsymbol{a}$, and then acting according to policy $\pi$:
\begin{equation} \label{Q_1}
Q^\pi(\boldsymbol{s},\boldsymbol{a})={\mathrm{E}}_{\tau \sim \pi} \left[\sum_{t=0}^{\infty} \gamma^{t} r_{t} \mid \boldsymbol{s}_{0}=s, \boldsymbol{a}_{0}=a\right].
\end{equation}


Then the target of DRL is finding optimal policy $\pi^{*}$ to maximize the state-action value function,
\begin{equation}
\pi^{*}=\arg \max _{\boldsymbol{a} \sim \pi} Q^\pi(\boldsymbol{s},\boldsymbol{a}).
\end{equation}

To address the high dimension optimization problem, deep reinforcement learning utilizes the neural network to approximate the state-action function and policy.

For the single-period optimization task, like IB-VVC, the optimization objective is to maximise the recent results and not consider the future. Single-period DRL is a specific approach for these tasks. Compared with the general DRL algorithms, single-period DRL would improve the VVC performance by reducing the learning difficulties of DRL and avoiding the overestimation issues of Q networks \cite{liu2022OSTC}.
The objective of single-period DRL is to find a policy to maximize the recent reward. Correspondingly, the state action function (also named as critic) 
$Q^\pi(\boldsymbol{s},\boldsymbol{a})$ is:
\begin{equation}\label{one_Q_1}
Q^\pi(\boldsymbol{s},\boldsymbol{a})={\mathrm{E}}\left[ r \mid \boldsymbol{s}_{0}=\boldsymbol{s}, \boldsymbol{a}_{0}=\boldsymbol{a}\right].
\end{equation}
Noting that setting $\gamma = 0$ of Equations \eqref{Q_1} derives the Equations \eqref{one_Q_1} directly. 

\subsection{Formulating IB-VVC as an MDP}


DRL learns from the MDP data.
The main components of the MDP for IB-VVC are as follows:
\begin{itemize}
\item[1)] State: $\boldsymbol{s} = (\boldsymbol{P}, \boldsymbol{Q}, \boldsymbol{V}, \boldsymbol{Q}_{C})$, where $\boldsymbol{P}, \boldsymbol{Q}, \boldsymbol{V}, \boldsymbol{Q}_{C}$ are the vector of active, reactive power injection, the voltage of all buses, and the reactive power outputs of  controllable IB-ERs and SVCs.
Similar to \cite{haotian_Two_Stage,haiyueRobust}, the topology information is assumed to not change in the control process, so the state can omit the topology information for simplicity.

\item[2)] Action: The action $ \boldsymbol{a}= \boldsymbol{Q}_{C}$, where $\boldsymbol{Q}_{C}$ is the reactive power outputs of all IB-ERs and SVCs.  The range of IB-ERs is  $\left|\boldsymbol{Q}_{G}\right| \leq \sqrt{\boldsymbol{S}_{G}^{2}-\overline{\boldsymbol{P}_{G }}^{2}}$, where $\overline{\boldsymbol{P}_{G}}$ is the upper limit of active power generation  \cite{Two_Timescale_yang,Stochastic_Kekatos}. The range of SVCs is $\underline{\boldsymbol{Q}_{C }} \leq \boldsymbol{Q}_{C} \leq \overline{\boldsymbol{Q}_{C}}$, where $\overline{\boldsymbol{Q}_{C}}$ and $\underline{\boldsymbol{Q}_{C}}$ is the upper and bottom limit of reactive power generation.

\item[3)] Reward: 
The reward for power loss $r_{p}$ is the negative of active power loss 
 \begin{equation}
  r_{p} =  - \sum_{i=0}^n  P_i,
 \end{equation}
and the reward of voltage violation rate $r_v$ is  
\begin{equation}
r_v= -\sum_{i=0}^n \left[\max \left(V_i-\bar{V}, 0\right)+\max \left(\underline{V}-V_i, 0\right)\right].
\end{equation}
The overall reward is 
\begin{equation}
    r = r_p +c_v r_v,
\end{equation} 
where $c_v$ is the weight ratio.
\end{itemize}

\section{Impact of Large Action Space on Learning Difficulties}\label{Proposition}

DRL algorithms learn to make an optimal decision by trial and error in the action space.
Intuitively, the larger action space, the more difficulties of DRL to find the optimal action.
In this section, {\color{red}we first analyze the impact of large action space,}
and then give five reasons that a large action space increases the learning difficulties of actor-critic algorithms.
Since off-policy actor-critic algorithms, such as DDPG\cite{lillicrap2016continuous}, TD3\cite{fujimoto2018addressing} and SAC\cite{haarnoja2018soft}, are popular for IB-VVC, this section takes off-policy actor-critic algorithms as an example to analyze the problems.

\subsection{The Impact of Large Action Space}

\begin{figure}[!t] 
\centering
\includegraphics[width=3.5in]{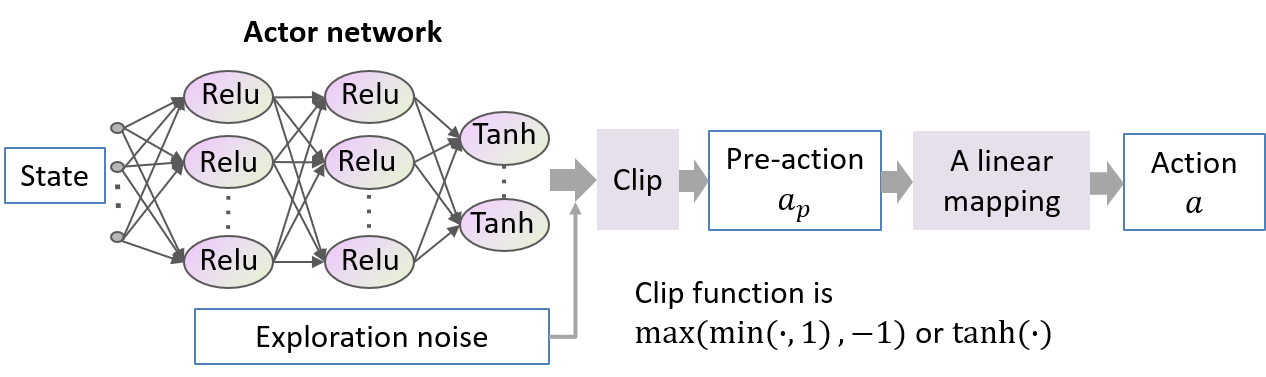}
\caption{The process of generating actions in the training stage.}
\label{actor_process}
\end{figure}

For IB-VVC, the action space $\mathbb{R}_a$ is a bounded box, while the optimal action $\boldsymbol{a}^*$  is only a curve. {\color{blue} Compared with a curve, the action space is considerably large.
To investigate the influence of large action space, we assume we can change the action space. In the process of enlarging or shrinking the action space, the optimal action curve is not changed (no active constraints for control variables).}

The process of generating actions by actor network is shown in Fig. \ref{actor_process}.
To satisfy the box constraints of the action, the activation function of the output layer of the actor network is $\tanh$.
After adding the exploration noise, the output vector is transformed into the pre-action $\boldsymbol{a}_{p}$ by the clip function. The clip function is $\max(\min(\cdot,1), -1)$ for DDPG and TD3, and $\tanh$ for SAC.
Recently, DRL algorithms focus more on the training performance and the convenience of training actor networks, and the pre-action space is an open interval $(-1,1)$ or closed interval $[-1,1]$ may be not considered rigorously. For convenience, we do not distinguish the open or closed interval, and just use the open interval $(-1,1)$ in this paper.
DRL algorithms generally store pre-actions $\boldsymbol{a}_{p}$ in data buffer rather than actions $\boldsymbol{a}$. The pre-actions $\boldsymbol{a}_{p}$ can be seen as the normalizing value of actions. Empirically, learning from normalizing data may have better results.

The pre-action $\boldsymbol{a}_{p}$ in the pre-action space $(-1,1)$ is transformed into the action space $(\bar{\boldsymbol{a}}, \underline{\boldsymbol{a}})$  through a linear mapping:
\begin{equation}\label{linear_map}
\begin{split}
\boldsymbol{k} &= (\bar{\boldsymbol{a}} - \underline{\boldsymbol{a}})/ 2 \\
\boldsymbol{b} &= (\bar{\boldsymbol{a}} + \underline{\boldsymbol{a}})/ 2 \\
\boldsymbol{a} &= \boldsymbol{k} \boldsymbol{a}_{p} + \boldsymbol{b}.
\end{split}
\end{equation}
The action $\boldsymbol{a}$ would be executed to the environment.

\begin{figure}[!t] 
\centering
\includegraphics[width=3in]{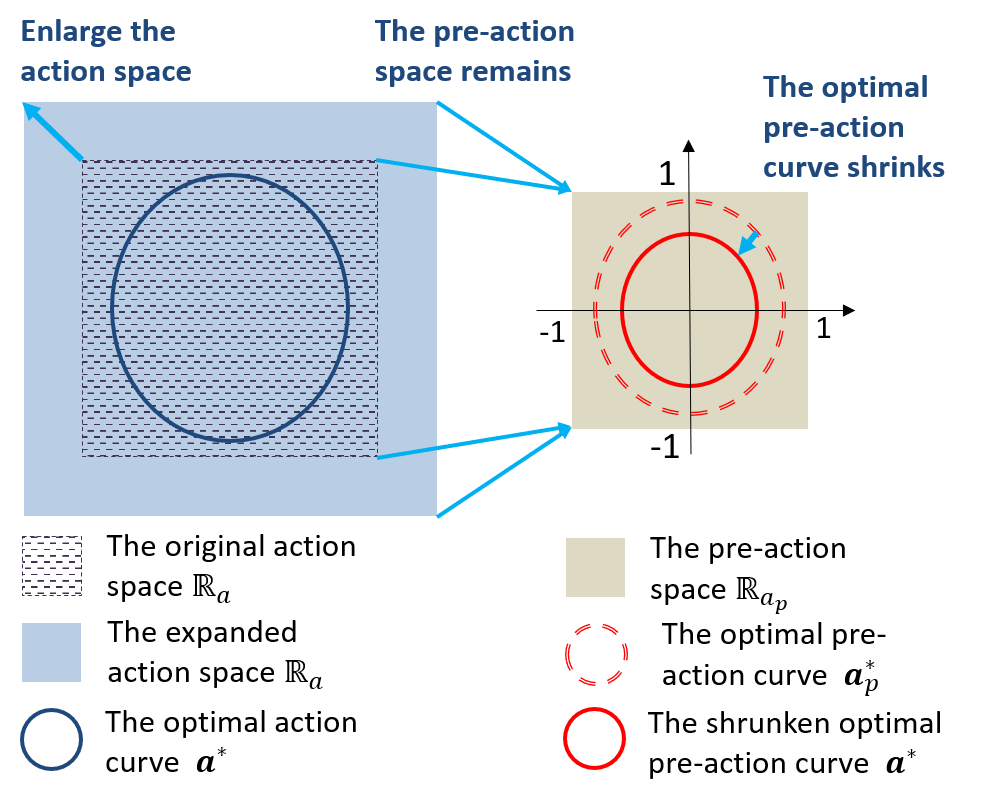}
\caption{The change of the pre-action space  $\mathbb{R}_{a_p}$ and the optimal pre-action $\boldsymbol{a}_p^*$ after increasing action space $\mathbb{R}_a$. Increase the action space $\mathbb{R}_a$, the optimal action curve $\boldsymbol{a}^*$ remains, the pre-action space $\mathbb{R}_{a_p}$ remains, the optimal pre-action curve $\boldsymbol{a}_p^*$ shrinks.}
\label{action_space}
\end{figure}

Fig. \ref{action_space} shows the change of the pre-action space and the optimal pre-action after enlarging the action space.
If the optimal action is in the action space and no active action constraint, after enlarging the action space $\mathbb{R}_a$, the optimal action curve $\boldsymbol{a}^*$ is not changed, and the pre-action space $\mathbb{R}_{a_p}$is always $(-1,1)$. According to the equations \eqref{linear_map}, the action space increases, $\bar{\boldsymbol{a}}-\underline{\boldsymbol{a}}$ increases, and $\boldsymbol{k}$ increases. The optimal pre-action satisfies $\boldsymbol{a}_p^* = (\boldsymbol{a}^* - \boldsymbol{b})/\boldsymbol{k}$, $\boldsymbol{k}$ increases, and  $\boldsymbol{a}_p^*$ shrinks.



\subsection{Five Reasons for Learning Difficulties due to Large Action Space}

The learning process of DRL can be roughly divided into two sub-processes: generating data and training neural networks.
The reasons for large action space increasing learning difficulties of DRL algorithms also occur in the two sub-processes.


In the process of generating data:
\begin{itemize}
\item[(1)] The large action space \textbf{exaggerates exploration noise} by the linear mapping process, thus leading to worse VVC performance in the training stage.
For a large action space, more pre-actions are far away from the optimal pre-actions.
\end{itemize}

In the process of training neural networks:
\begin{itemize}
\item[(2)] Large action space \textbf{leads to a large critic error}. The critic learning from the data farther away from the optimal data has a higher generalization error for optimal data. The larger action space, the more data farther away from the optimal data tend to be generated. The problem is significant in the free exploration stage and gradually decreases in the learning stage.
\item[(3)] \textbf{The large critic error would propagate to the actor network}, which may lead to a large actor error.
\item[(4)] Large action space may \textbf{exaggerate the actor error mapping from pre-actions to actions}.
When the action space increase from $(\underline{\boldsymbol{a}}, \bar{\boldsymbol{a}})$ to $\lambda(\underline{\boldsymbol{a}}, \bar{\boldsymbol{a}})$, where $\lambda>1$, the coefficient $\boldsymbol{k}$ will increase to $\lambda \boldsymbol{k}$ in Equation \eqref{linear_map}.
The action error $\epsilon_a$ maybe be exaggerated to $\lambda \boldsymbol{k} \epsilon_{a}$ in action space.



\item[(5)] \textbf{The coupling of actor and critic slows down the convergence of DRL further.}
The inaccurate critic cannot provide efficient guidance for the actor network. The inaccurate actor network also cannot provide optimal actions for updating the critic network.
The problem is significant in the initial learning stage that the random data dominate in the data buffer. Both the critic and actor cannot learn a good performance from those random data.  The DRL agent only can generate slightly better actions. The improvement rate depends on the generalization capabilities of the actor  and critic networks.
\end{itemize}

\begin{figure}[!t] 
\centering
\includegraphics[width=3.5in]{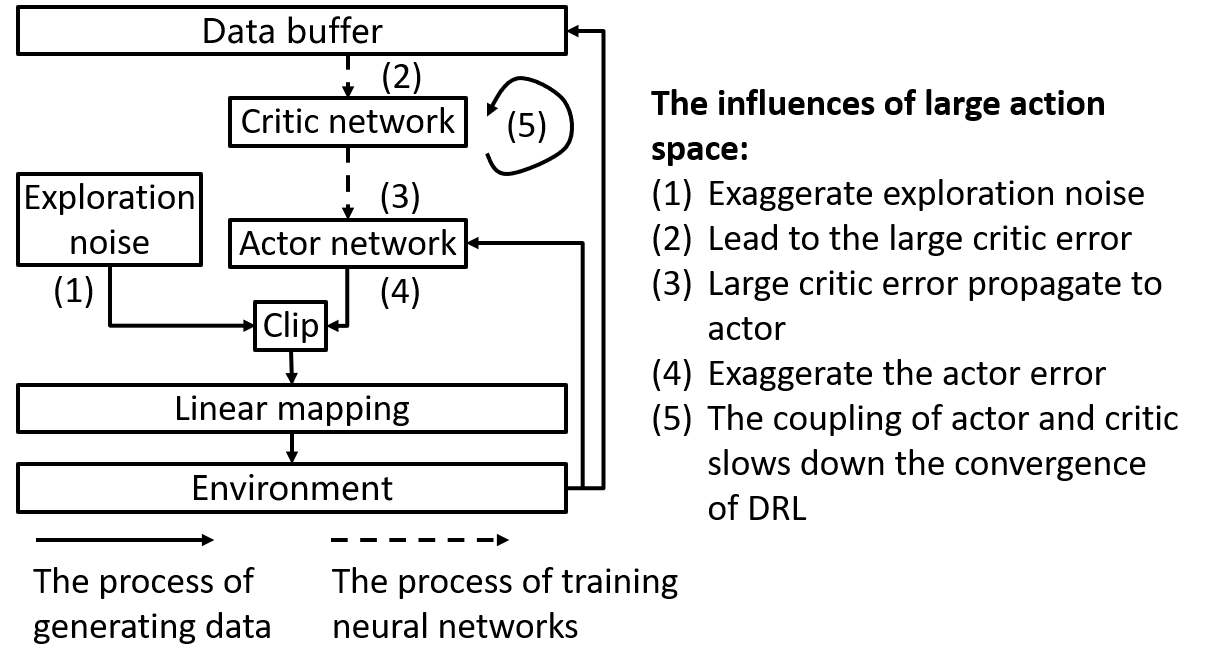}
\caption{Five reasons for learning difficulties due to large action space}
\label{5reason}
\end{figure}

\section{Reference-Model-Based Optimization for Inverter-based Volt-Var Control}

Most of time, the accurate power flow model of distribution networks is difficult to acquire, whereas the model-based optimization under the inaccurate model still has a good reference.
In this section, we introduce reference-model-based optimization.  It is similar to the existing model-based optimization method, and the difference is it utilizes a reference model rather than the accurate model. A reference model is not just an inaccurate model, it needs to satisfy more conditions. 



\begin{definition} \label{reference-model}
For a specific optimization task, $\boldsymbol{a}_m$ is the optimal action under the inaccurate model, $\boldsymbol{a}^*$ is the optimal action under the accurate model, and $\boldsymbol{a}_r^*$ is the optimal residual actions that $\boldsymbol{a}_r^*  = \boldsymbol{a}^* -\boldsymbol{a}_m$.
If the optimal residual action  $\boldsymbol{a}_r^*$ satisfy $0< {\|\boldsymbol{a}_r^*\|}_2 < {\|\boldsymbol{a}^*\|}_2 $, then the inaccurate model is the reference model for the specific optimization task. The optimization task is a reference-model-based optimization, and $\boldsymbol{a}_m$ is the reference action.
\end{definition}

The definition of the reference model is motivated by the large action problem. Reference-model-assisted DRL can select a small action space by learning the optimal residual action $\boldsymbol{a}_r^*$. It is a necessary condition for the effectiveness of reference-model-assisted DRL.



Generally, accurate power flow models are difficult to obtain, whereas inaccurate models often satisfy the condition of a reference model.
For example, the resistance and reactance parameters usually are the nominal values, whereas those parameters are influenced by environmental conditions such as temperature and humidity which are seldom considered. Sometimes, the susceptance is also omitted in power flow models.
The optimal decisions based on the inaccurate power flow model tend to be not optimal for the real ADN, which may have a slightly worse VVC performance. However, empirically, the non-optimal decisions still work and achieve a decent VVC performance. 
This may indicate that  those power flow models is the reference model for the optimization task, and $\boldsymbol{a}_r^*$ is less than  $\boldsymbol{a}^*$ considerably.






Since the accurate power flow model is difficult to acquire, it is not easy to verify whether the obtained nominal model is a reference model from the definition \ref{reference-model}. The non-linearity of IB-VVC problems hinders the verification further. 
However, we can  verify it from the simulation results empirically. If reference-model-assisted DRL has a better VVC performance than both model-based optimization with an inaccurate model and the general DRL, we can say reference-model-assisted DRL have advantages. The inaccurate model is more likely to satisfy the definition of a reference model.


\section{Reference-Model-Assisted Deep Reinforcement Learning}

Large action space increases the learning difficulties of DRL. To decrease the action space of DRL, a reference-model-assisted DRL approach is proposed in this section.
Firstly, we give the introduction to the reference-model-assisted DRL approach.
The reference-model-assisted DRL approach cooperates well with any policy gradient or actor-critic DRL algorithms for continuous action problems. 
Secondly, we take the SAC algorithm as an example and design a practical reference-model-assisted DRL algorithm.


\subsection{The Framework of Reference-Model-Assisted Deep Reinforcement Learning}\label{subsection_WMA_SAC}

\begin{figure}[!t] 
\centering
\includegraphics[width=3.2in]{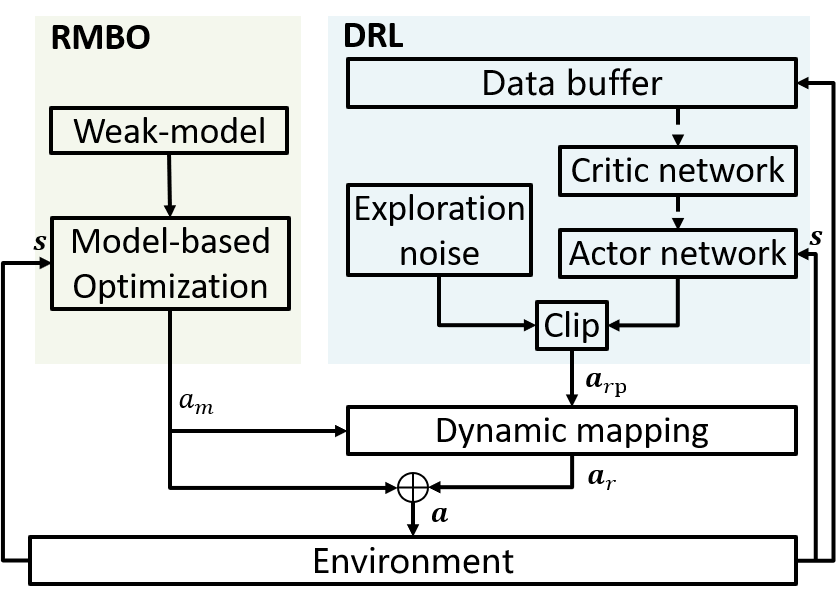}
\caption{The framework of the reference-model-assisted DRL.}
\label{WMA-DRL}
\end{figure}

Fig. \ref{WMA-DRL} shows the framework of reference-model-assisted DRL. The objective of reference-model-assisted DRL is to learn the optimal residual action $\boldsymbol{a}_r^*$ from the residual action space $\mathbb{R}_{\boldsymbol{a}_r^*}$.
For reference model, $0<\|\boldsymbol{a}_r^*\|<\|\boldsymbol{a}^*\|$, and the residual action space $\mathbb{R}_{\boldsymbol{a}_r^*}$ is less than the action space $\mathbb{R}_{\boldsymbol{a}}$.
The reference-model-assisted DRL can select a smaller action space $\mathbb{R}_{\boldsymbol{a}_r^*}$ than the general DRL method of learning the optimal action $\boldsymbol{a}^*$ directly in the action space $\mathbb{R}_{\boldsymbol{a}}$.
According to the proposition of Section III, it reduces the learning difficulties of DRL, thus having a faster convergence and optimal results. 

\begin{figure}[!t]
\centering
\subfigure[``too small" residual action space]{\begin{minipage}[t]{0.4\linewidth}\label{cannot_reach_optimal}
    \includegraphics[width=1.3in]{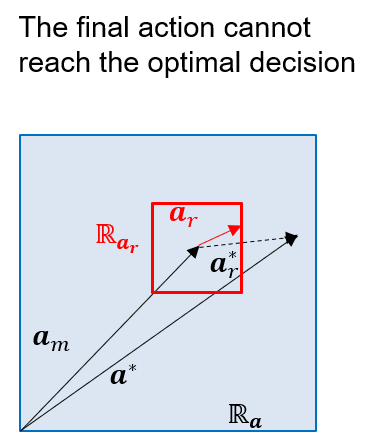}
    \end{minipage}
    }
\subfigure[``too large" residual action space]{
\begin{minipage}[t]{0.4\linewidth}\label{exceed_constraints}
\includegraphics[width=1.3in]{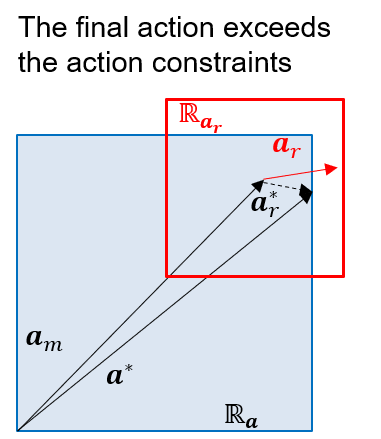}
\end{minipage}
    }
\caption{The problem of ``too small" or ``too large"  residual action space.}
\end{figure}

{\color{red}Designing the residual action space is a key component for reference-assisted DRL.}
Since both the optimal action $\boldsymbol{a}^*$ and the optimal residual action $\boldsymbol{a}^*_r$ are unknown beforehand, we can roughly set the residual action space of reference-model-assisted DRL smaller than the original action space such that $\mathbb{R}_{\boldsymbol{a}_r} = (-\boldsymbol{\delta},\boldsymbol{\delta})$, where $\boldsymbol{\delta} < ( \bar{\boldsymbol{a}} - \underline{\boldsymbol{a}})/2$.
Then the residual pre-action $\boldsymbol{a}_{rp}$ outputted by the actor network in the space $(-1,1)$ are mapped into the residual action space $(-\boldsymbol{\delta},\boldsymbol{\delta})$.
Since the optimal action cannot be known beforehand, two problems would appear. If we set a too small residual action space for reference-model-assisted DRL, it would lead to the final action cannot reach the optimal action, as shown in Fig. \ref{cannot_reach_optimal}.  If we set a ``too large" residual action space, it may lead to the final action exceeding the action constraints, as shown in Fig. \ref{exceed_constraints}. 

For the first problem, reference-model-assisted DRL under a ``too-small" residual action space still can have a better performance than the reference-model-based optimization results. As shown in Fig. \ref{cannot_reach_optimal}, the final action $\boldsymbol{a} = \boldsymbol{a}_m + a_r$ are closer to the optimal action $\boldsymbol{a}^*$ compared with the reference model optimization action $\boldsymbol{a}_m$.


To address the second problem, we propose a dynamical mapping trick to adjust the residual action space $\mathbb{R}_{\boldsymbol{a}_r}$ into the action space $(\underline{\boldsymbol{a}}, \bar{\boldsymbol{a}})$,
where
\begin{equation}\label{clip_action_space}
\begin{split}
\underline{\boldsymbol{\delta}}_{a_r} &= -\boldsymbol{\delta} - \min(\boldsymbol{a}_m -\boldsymbol{\delta} -\underline{\boldsymbol{a}}, 0) \\
\bar{\boldsymbol{\delta}}_{a_r} &= \boldsymbol{\delta} - \max(\boldsymbol{a}_m + \boldsymbol{\delta} -\bar{\boldsymbol{a}}, 0)
\end{split}
\end{equation}
It guarantees the final action always in the action space $[\underline{\boldsymbol{a}} ,\bar{\boldsymbol{a}}]$. 

Then, the dynamical mapping trick linearly maps the residual pre-action $\boldsymbol{a}_{rp}$ of reference-model-assisted DRL to the residual action space $\mathbb{R}_{\boldsymbol{a}_r}$,
\begin{equation}\label{map_clip_action_space}
\begin{split}
\boldsymbol{k} &= (\bar{\boldsymbol{\delta}}_{a_r} - \underline{\boldsymbol{\delta}}_{a_r})/ 2 \\
\boldsymbol{b} &= (\bar{\boldsymbol{\delta}}_{a_r} + \underline{\boldsymbol{\delta}}_{a_r})/ 2 \\
\boldsymbol{a}_r &= \boldsymbol{k} \boldsymbol{a}_{rp} + \boldsymbol{b}
\end{split}
\end{equation}

As shown in Fig \ref{WMA-DRL}, for the state $\boldsymbol{s}$ of each step, the model-based optimization method calculates the reference action $\boldsymbol{a}_m$ based on the reference model, the actor network of the reference-model-assisted DRL output the residual pre-action $\boldsymbol{a}_{rp}$, and then the dynamic mapping trick transforms the residual pre-action $\boldsymbol{a}_{rp}$ to the residual action $\boldsymbol{a}_r$. The final action is $\boldsymbol{a} = \boldsymbol{a}_m + a_r$. After applying the final action $\boldsymbol{a}$ to the environment, the reference-model-assisted DRL observes a reward $r$ and a new  state $\boldsymbol{s}^{\prime}$. 


The state-action function $Q^{\pi}(\boldsymbol{s}, \boldsymbol{a}_m, \boldsymbol{a}_{rp})$ for reference-model-assisted DRL is:
\begin{equation}\label{Q}
\begin{split}
Q^{\pi_m, \pi_{rp}}(\boldsymbol{s},\boldsymbol{a}_m, \boldsymbol{a}_{rp})={\mathrm{E}}_{\tau \sim \pi_m, \pi_{rp}} \Big[ \sum_{t=0}^{\infty} \gamma^{t} r_{t} \mid \boldsymbol{s}_{0}=\boldsymbol{s},  \\
\boldsymbol{a}_{w0}=\boldsymbol{a}_{m}, \boldsymbol{a}_{r0}=\boldsymbol{a}_{rp} \Big].
\end{split}
\end{equation}
where $\pi_m$ is the reference-model-based optimization policy and $\pi_{rp}$ is the reference-model-assisted DRL policy.


For single-period optimization tasks, like IB-VVC, $\gamma = 0$ and the single-period state-action function is 
\begin{equation}\label{one_Q}
Q^{\pi_m, \pi_{rp}}(s,\boldsymbol{a}_{m}, \boldsymbol{a}_{rp})={\mathrm{E}}\left[ r(s,\boldsymbol{a}_m, \boldsymbol{a}_{rp}) \right],
\end{equation}

The target of DRL is finding optimal policy $\pi_{rp}^*$ to maximize state-action value function,
\begin{equation}\label{gradient_actor}
\pi_{rp}^{*}=\arg \max _{\boldsymbol{a}_{rp} \sim \pi_{rp}} Q^{\pi_m, \pi_{rp}}(\boldsymbol{s}, \boldsymbol{a}_m, \boldsymbol{a}_{rp}).
\end{equation}

This paper considers the deterministic optimization problem. The model-based optimization policy $\boldsymbol{a}_{m}$ is deterministic, and the pair of a state $\boldsymbol{s}$ and a reference action $\boldsymbol{a}_m$ is not changed. This means for the same $\boldsymbol{s}$ and $\boldsymbol{a}_{rp}$, $Q^{\pi_m, \pi_{rp}}(\boldsymbol{s},\boldsymbol{a}_{m}, \boldsymbol{a}_{rp}) = Q^{\pi_{rp}}(\boldsymbol{s},\boldsymbol{a}_{rp})$, so we can omit $\boldsymbol{a}_m$ in the state-action function. 
In function \eqref{gradient_actor}, $\boldsymbol{a}_m$ can be omitted as well. In reply buffer, we only need to store the data $(\boldsymbol{s},\boldsymbol{a}_{rp}, r, \boldsymbol{s}^\prime)$.



\subsection{A Practical Algorithm: Reference-Model-Assisted Soft Actor-Critic}

\begin{algorithm}[!t]
  \caption{Reference-Model Assisted Soft Actor-Critic Algorithm} \label{reference-model-assited-algorihtm}
  \begin{algorithmic}[1]
  \Require
  Initial policy parameters $\boldsymbol{\theta}$, Q-function parameters $\boldsymbol{\phi}_1$, $\boldsymbol{\phi}_2$,  empty replay buffer $\mathcal{D}$.
  \Require Reference power flow model.
  \Require Set the value $\boldsymbol{\delta}$ for residual action space $\mathbb{R}_{\boldsymbol{a}_r}$.
\For {each environment step}

\State  Observe state $\boldsymbol{s}$, calculate the reference action $\boldsymbol{a}_m$ by 
\Statex \quad \ \  calling Pandapower based on the reference model,
\Statex \quad \ \ and calculate the DRL action  $\boldsymbol{a}_{rp} \sim \pi_{rp}^{\boldsymbol{\theta}}(\cdot \mid \boldsymbol{s})$.
\State  Obtain the residual action $\boldsymbol{a}_r$ by mapping the residual 
\Statex \quad \ \  pre-action $\boldsymbol{a}_{rp}$ to the action $\boldsymbol{a}_r$ using \eqref{map_clip_action_space}.
\State The final action $ \boldsymbol{a} = \boldsymbol{a}_m + \boldsymbol{a}_r$, execute $\boldsymbol{a}$ in the
\Statex \quad \ \ environment, and observe reward $r$ and next state $s^\prime$.
\State Store $(\boldsymbol{s},\boldsymbol{a}_{rp},r)$ in replay buffer.
\If {it's time to update} 

\For {$j$ in range (how many updates)} 

 \State Randomly sample a batch of transitions
\Statex \qquad \qquad \  $B={(\boldsymbol{s},\boldsymbol{a}_{rp},r, \boldsymbol{s}^\prime, d)}$ from $\mathcal{D}$.
 
\State Update $Q_{\boldsymbol{\phi}_1}$ and $Q_{\boldsymbol{\phi}_2}$ by minimizing the loss 
\Statex \qquad \qquad \ \ function \eqref{L_one_step_Q} for   single multi-period 
\Statex \qquad \qquad \ \ optimization problems,
 or  \eqref{L_critic} for  multi-period 
  \Statex \qquad \qquad \ \  optimization problems.
\State Update $\pi_{rp}^{\boldsymbol{\theta}}$ by maximizing the loss function 
\Statex \qquad \qquad \ \ \eqref{loss_actor}.
\State Update $\alpha$ by minimizing the loss function \eqref{entropy_adjust}.
\EndFor
\EndIf
\EndFor
\end{algorithmic}
\end{algorithm}

The reference-model-assisted DRL approach is compatible with any policy gradient or actor-critic DRL algorithms for continuous action space. This paper takes soft actor-critic (SAC) as an example and designs  reference-model-assisted SAC. SAC includes the actor network (policy) and the critic network (state-action value function).
SAC has four critical technologies.   The first two technologies,``replay buffers" and ``target networks", inherited from DDPG \cite{lillicrap2016continuous} improve its stability. The third technology ``clipped double-Q Learning" inherited from TD3 \cite{fujimoto2018addressing} eliminates the overestimation problem. The fourth technology  "entropy regularization" provides a more stable Q value estimation and prevents the policy from prematurely converging to a bad local optimum.
In reference-model-assisted SAC, the critic network and actor network are the same as the original SAC except for replacing action with the residual pre-action $\boldsymbol{a}_{rp}$. 



The entropy-regularized critic is 
\begin{equation}
\begin{split}
 Q^{\pi_{rp}}(\boldsymbol{s}, \boldsymbol{a}_{rp})=&{\mathbb{E}}_{\tau \sim \pi_{rp}} \Big[\sum_{t=0}^{\infty} \gamma^{t} r_{t}+\alpha \sum_{t=1}^{\infty} \gamma^{t} H\left(\pi\left(\cdot \mid \boldsymbol{s}_{t}\right)\right) \\ 
 \mid \boldsymbol{s}_{0}=\boldsymbol{s}, 
 & \boldsymbol{a}_{r0}=\boldsymbol{a}_{rp} \Big], 
 \end{split}
\end{equation}
where $H\left(\pi_{rp}\left(\cdot \mid \boldsymbol{s}_{t}\right)\right)=\underset{\boldsymbol{a} \sim \pi\left(\cdot \mid \boldsymbol{s}_{t}\right)}{\mathbb{E}}\left[-\log \pi_{rp}\left(\cdot \mid \boldsymbol{s}_{t}\right)\right]$ is the entropy of the stochastic policy at $\boldsymbol{s}_{t}$, $\alpha$ is the temperature parameter. $Q^{\pi_{rp}}(\boldsymbol{s}, \boldsymbol{a}_{rp})$ satisfies the  Bellman equation, 
\begin{equation}
\begin{aligned}
Q^{\pi_{rp}}(\boldsymbol{s}, \boldsymbol{a}_{rp}) &=\underset{\substack{\boldsymbol{s}^{\prime} \sim P \\
\boldsymbol{a}_{rp}^{\prime} \sim \pi}}{\mathrm{E}}\left[r+\gamma\left(Q^{\pi_{rp}}\left(\boldsymbol{s}^{\prime}, \boldsymbol{a}_{rp}^\prime\right)+\alpha H\left(\pi_{rp}\left(\cdot \mid \boldsymbol{s}^{\prime}\right)\right)\right)\right].
\end{aligned}
\end{equation}

The same as the SAC, the reference-model-assisted SAC uses ``clip double Q learning" and ``target networks" to improve performance. Two critic neural network $Q_{\boldsymbol{\phi}_1}, Q_{\boldsymbol{\phi}_2}$ and two target critic neural networks $Q_{\boldsymbol{\phi}_{targ,1}}, Q_{\boldsymbol{\phi}_{targ,2}}$ are utilized. $\boldsymbol{\phi}_1,\boldsymbol{\phi}_2,\boldsymbol{\phi}_{targ,1}, \boldsymbol{\phi}_{targ,2} $ are the parameters of critic networks. The critic networks are learned by minimizing the mean-squared Bellman error $L_{\boldsymbol{\phi}_{i}}$,
\begin{equation}\label{L_critic}
 L_{\boldsymbol{\phi}_{i}}=\frac{1}{|B|}\sum_{\left(\boldsymbol{s}, \boldsymbol{a}_{rp}, r, \boldsymbol{s}^{\prime}, d\right) \sim \mathcal{D}}\left[\left(Q_{\boldsymbol{\phi}_{i}}(\boldsymbol{s}, \boldsymbol{a}_{rp})- y \right)^{2}\right],  
\end{equation}
where $i \in 1,2$, and the target $y$ is 
\begin{equation}
\begin{split}
    y =& r \! + \! \gamma(1\!-\!d)\left(\min _{j=1,2} Q_{\boldsymbol{\phi}_{\text {targ }, j}}\left(\boldsymbol{s}^{\prime}, \tilde{\boldsymbol{a}}_{rp}^{\prime}\right)  \! - \! \alpha \log \pi_{rp}^{\boldsymbol{\theta}}\left(\tilde{\boldsymbol{a}}_{rp}^{\prime} \mid \boldsymbol{s}^{\prime}\right)\right)  , \\
    & \quad \tilde{\boldsymbol{a}}_{rp}^{\prime} \sim \pi_{rp}^{\boldsymbol{\theta}}\left(\cdot \mid \boldsymbol{s}^{\prime}\right),
    \end{split}
\end{equation}
where $\boldsymbol{\theta}$ is the parameter of actor network.

Considering the IB-VVC is a single-period optimization task, we adopt a single-period SAC from \cite{liu2022OSTC}. 
In single-period SAC, 
$\gamma = 0$  and the critic network is $Q^{\pi_{rp}}(\boldsymbol{s}, \boldsymbol{a}_{rp})=  {\mathbb{E}}_{\tau \sim \pi_{rp}}  (r)$. The critic networks are learned by minimizing the loss function 
\begin{equation} \label{L_one_step_Q}
     L_{\boldsymbol{\phi}_{i}}=\frac{1}{|B|}\sum_{ \boldsymbol{s} \sim \mathcal{D}}\left[\left(Q_{\boldsymbol{\phi}_{i}}(\boldsymbol{s}, \boldsymbol{a}_{rp})-r\right)^{2}\right]   
\end{equation}

The residual actor $\pi_{rp}^{\boldsymbol{\theta}}$ is designed as
\begin{equation}
 \pi_{rp}^{\boldsymbol{\theta}}(\boldsymbol{s})=\tanh \left(\mu_{\boldsymbol{\theta}}(\boldsymbol{s})+\sigma_{\boldsymbol{\theta}}(\boldsymbol{s}) \odot \xi\right), \quad \xi \sim \mathcal{N}(0, I),
\end{equation}
where $\boldsymbol{\theta}$ is the parameters of actor network, $\mu$ is the mean function, and $\sigma$ is the variance function.
The output of the actor $\pi_{rp}^{\boldsymbol{\theta}}$ is the residual pre-action.

We have the residual action $\boldsymbol{a}_r$ by processing the residual pre-action $\boldsymbol{a}_{rp}$ using the dynamic mapping trick \eqref{map_clip_action_space}. The final execution action is $\boldsymbol{a} = \boldsymbol{a}_m + \boldsymbol{a}_r$.

The actor $\pi_{rp}^{\boldsymbol{\theta}}$ is learned by maximizing the loss function $L_{\boldsymbol{\theta}}$,
\begin{equation}\label{loss_actor}
  L_{\boldsymbol{\theta}} =  \frac{1}{|B|}\sum_{s \sim \mathcal{D}} \left[\min _{j=1,2} Q_{\boldsymbol{\phi}_{j}}\left(\boldsymbol{s}, \pi_{rp}^{\boldsymbol{\theta}}(\boldsymbol{s})\right)-\alpha \log \pi_{rp}^{\boldsymbol{\theta}}\left( \cdot \mid \boldsymbol{s}\right)\right].
\end{equation}

The entropy regularization coefficient $\alpha$ can be a constant or adjustable dual variable by minimizing the loss function $ L(\alpha)$,
\begin{equation}\label{entropy_adjust}
    L(\alpha) =  \frac{1}{|B|} \sum_{\boldsymbol{s} \in B } [-\alpha \log \pi_{\boldsymbol{\theta}} (\boldsymbol{\cdot}|\boldsymbol{s}) - \alpha \mathcal{H}],
\end{equation}
where $\mathcal{H}$ is the entropy target. 

The detail of reference-model-assisted SAC is shown in Algorithm \ref{reference-model-assited-algorihtm}.

 \begin{table}[!t]
\renewcommand{\arraystretch}{1.3}
\caption{Hyper-parameters for the DRL algorithm \cite{liu2022OSTC}}
\label{DRL_paramter}
\centering
\begin{tabular}{ c c}
\hline
\text { Parameter } & \text { Value } \\
\hline 
\text { Optimizer } & \text { Adam } \\
\text { Activation function } & \text { ReLU } \\
 \text { Number of hidden layers } & 2  \\
 Actor hidden layer neurons & \{512, 512\} \\
 Critic hidden layer neurons & \{512, 512\} \\
 \text { Batch size } &  128 \\
 \text { Replay buffer size } & $ 3 \times 10^{4}$ \\
 \text {Critic learning rate} &  $ 3 \times 10^{-4}$\\
 \text {Actor learning rate   } &  $ 1 \times 10^{-4}$\\
 Voltage violation penalty $c_v$ & 50\\
 Initial random step & 960\\
 Iterations per time step & 4\\

Entropy target & $-\dim(\mathcal{A})$\\
Temperature  learning rate & $ 3 \times 10^{-4}$ \\
\hline 
\end{tabular}
\end{table}

\section{Simulation}

Numerical simulation was conducted on 33-bus, \cite{baran1989network}, 69-bus \cite{das2008optimal}, and 118-bus \cite{ZHANG_118} distribution networks to verify the two propositions: 1) \textcolor{red}{large} action space increases the learning difficulties of  DRL; and 2) reference-model-assisted DRL improves the VVC performance. The simulation environment setting was similar to our previous paper \cite{liu2022OSTC}.
In the 33-bus system, 3 IB-ERs were connected to bus 17, 21, and 24, and 1 SVC of 2 MVar was connected to bus 32.   
In the 69-bus system, 4 IB-ERs were connected to bus 5, 22, 44, 63, and 1 SVC was connected to bus 13.
In the 118-bus system, 8 IB-ERs were connected to bus 33, 50, 53, 68, 74, 97, 107, and 111,  and 2 SVC was connected to bus 44 and 104.
Each IB-ER had 1.5 MW active power and 2 MVar capacity. Each SVC had 2 MVar capacity.
All load and generation levels were multiplied with the fluctuation ratio \cite{haotian_Two_Stage} and a $20\%$ uniform distribution noise to reflect the variance.  
The normal voltage operation range for all buses was set to be [0.95, 1.05].
The algorithms were implemented in Python. The balanced power flow was solved by Pandapower \cite{pandapower2018} to simulate ADNs, and the implementation of the DRL algorithms used PyTorch. 
For easily to access the simulation setting and reproduce the results, the detailed simulation codes will be open source.

\begin{figure*}[ht]
\centering
\subfigure[33-bus]{\begin{minipage}[t]{0.315\linewidth}\label{wm1_large_sacle_33}
    \includegraphics[width=2.4in]{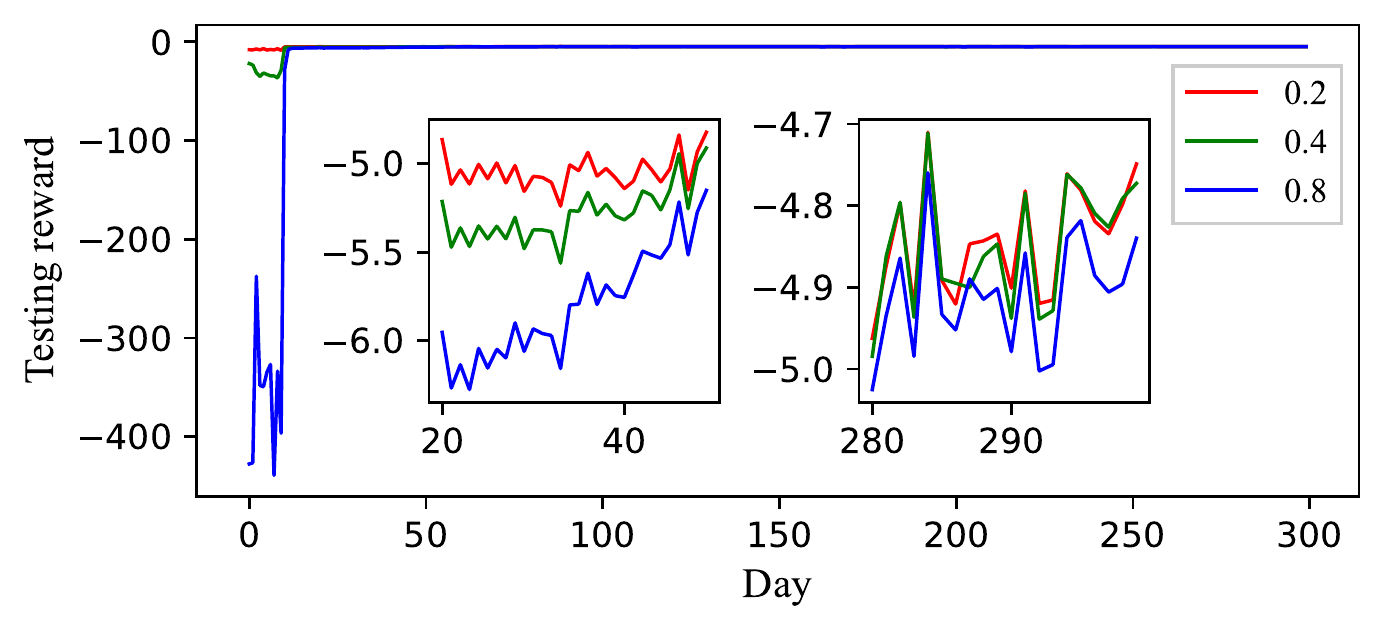}
    \end{minipage}
    }
\subfigure[69-bus]{
\begin{minipage}[t]{0.315\linewidth}\label{wm1_large_sacle_69}
\includegraphics[width=2.4in]{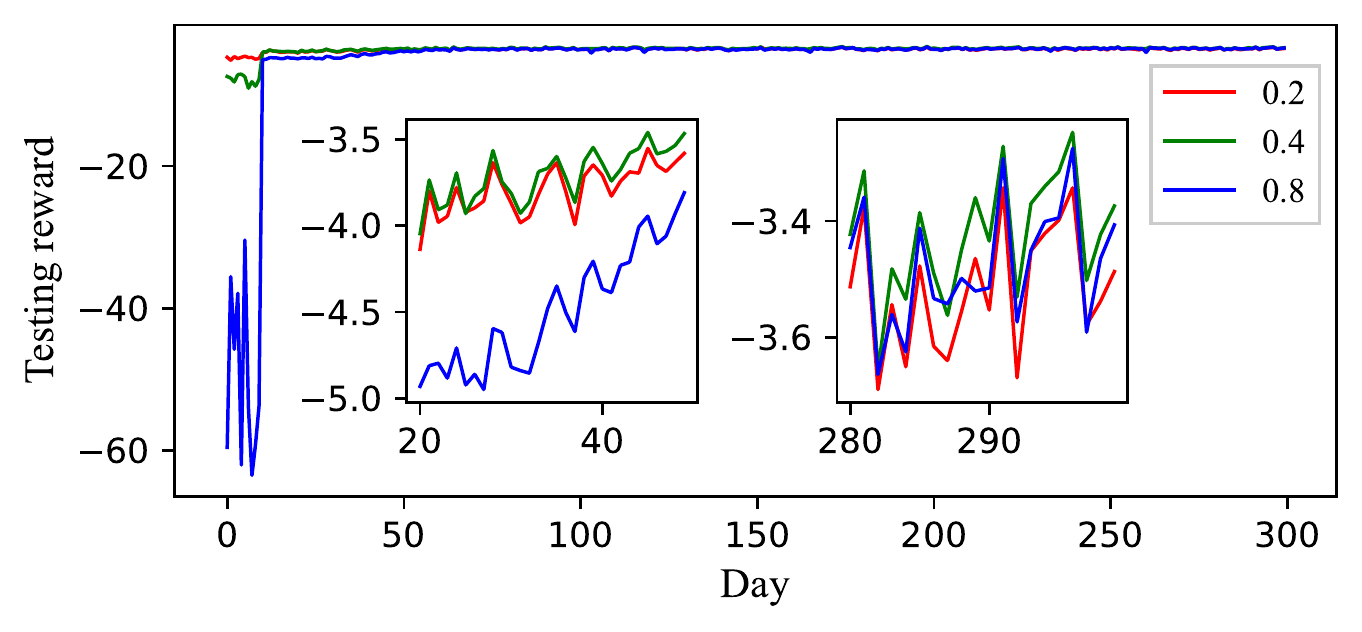}
\end{minipage}
    }
\subfigure[118-bus]{
\begin{minipage}[t]{0.315\linewidth}\label{wm1_large_sacle_118}
\includegraphics[width=2.4in]{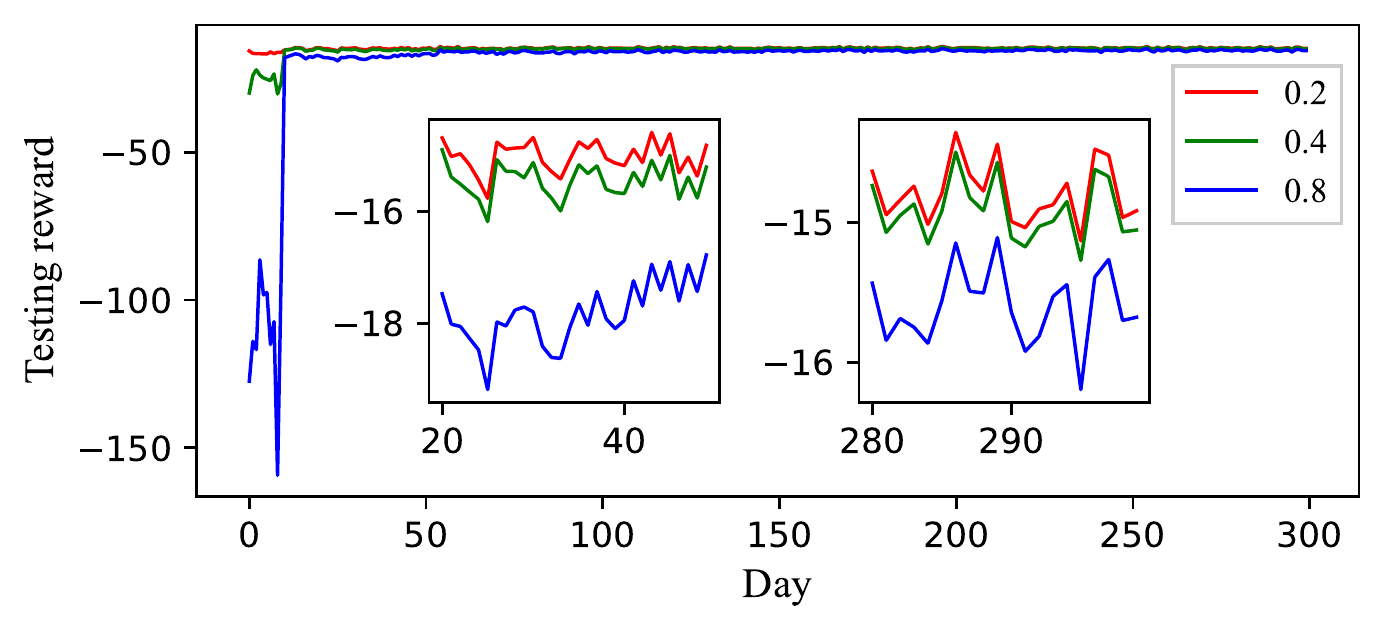}
\end{minipage}
    }
\caption{Testing results of the training stage for 33-bus and 69-bus test distribution networks. $\lambda = [0.2,0.4,0.8]$} \label{wm1_large_sacle}
\end{figure*}


\begin{figure*}[htb]
\centering


\subfigure[33-bus]{\begin{minipage}[t]{0.315\linewidth}\label{wm2_large_sacle_33}
    \includegraphics[width=2.4in]{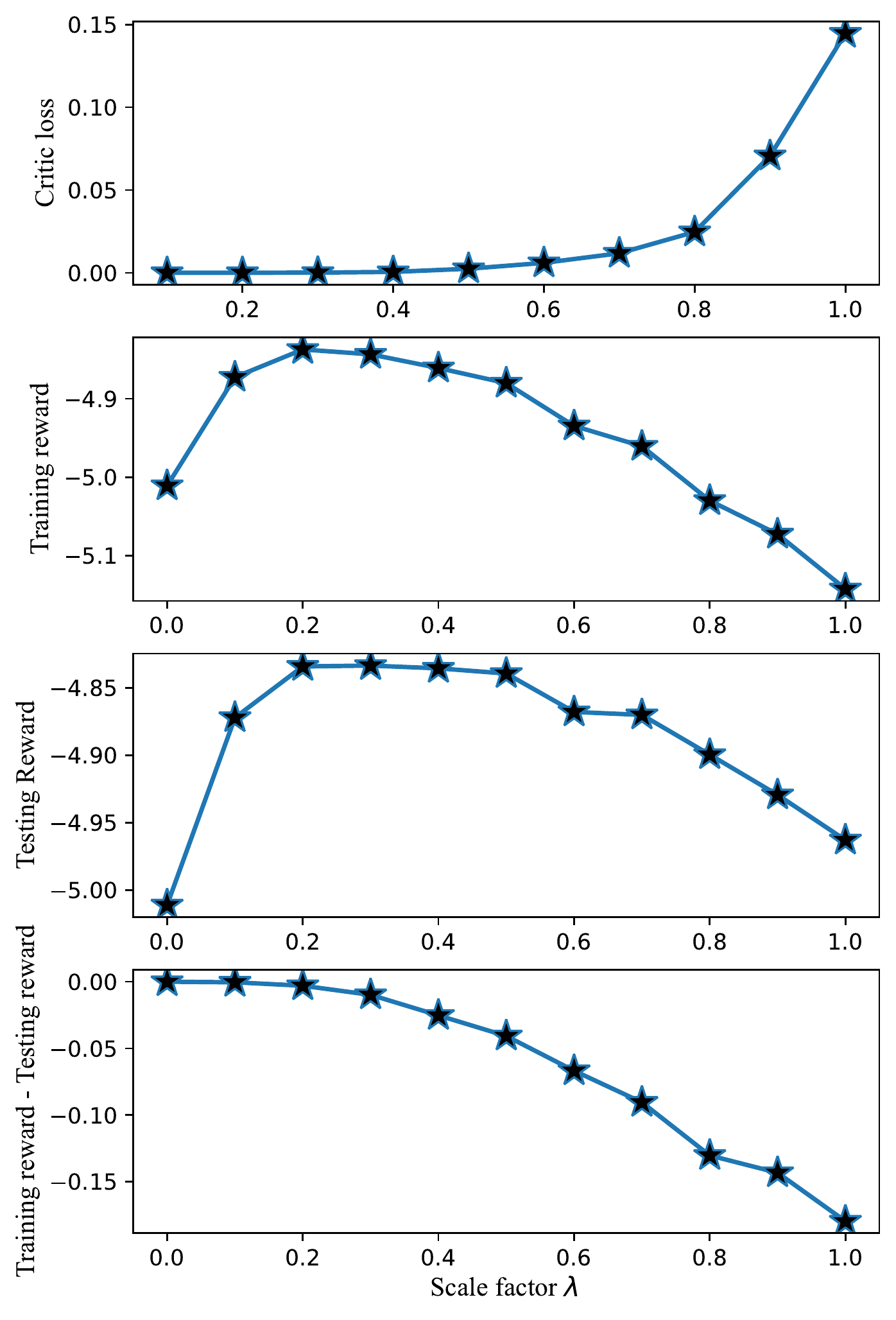}
    \end{minipage}
    }
\subfigure[69-bus]{
\begin{minipage}[t]{0.315\linewidth}\label{wm2_large_sacle_69}
\includegraphics[width=2.4in]{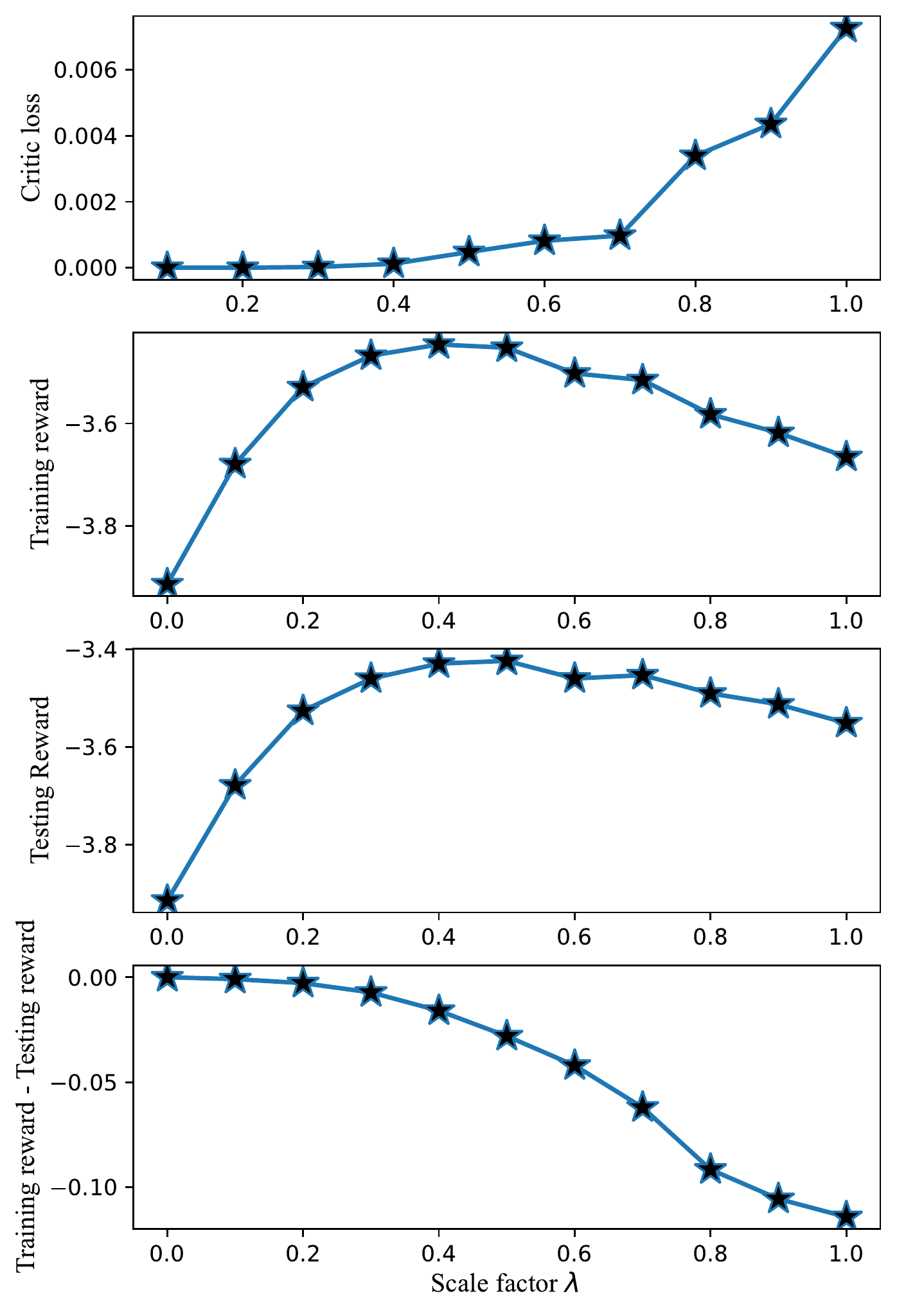}
\end{minipage}
    }
\subfigure[118-bus]{
\begin{minipage}[t]{0.315\linewidth}\label{wm2_large_sacle_118}
\includegraphics[width=2.4in]{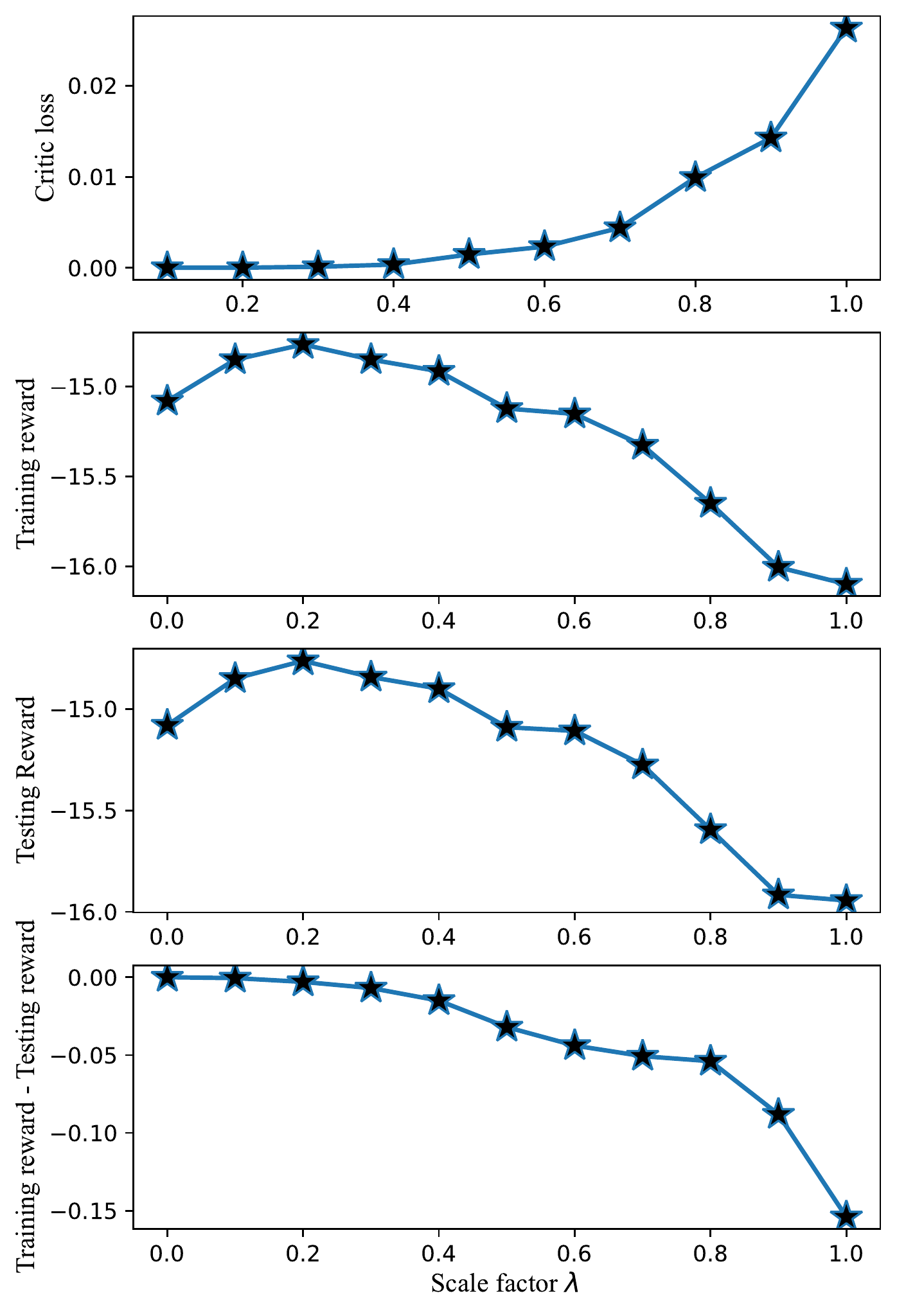}
\end{minipage}
    }
    
\caption{The change of the critic loss, the training reward, the testing reward, and the training reward - the testing reward with the increasing of residual action space in the final 50 days. The training reward - the testing reward is the  effect of exploration noise.}\label{wm2_large_sacle}
\end{figure*}

 We designed five classes of simulation experiments. 
\begin{itemize}
    \item[1)] \textbf{Model-based optimization under an accurate model:}
    Model-based optimization was solved by PandaPower with the interior point solver.
    We assumed the original parameters in test distribution networks are accurate.
    \item[2)] \textbf{Model-based optimization under a reference model:}  The same as the model-based optimization method expect that the inaccurate parameters of resistance and reactance of branches were 1.5 times of the original ones for 33-bus and 69 bus distribution networks, and 1.3 times for 118-bus distribution network.  
    \item[3)] \textbf{DRL:} Since IB-VVC is a single-period optimization problem, we used the single-period SAC adopted from the paper \cite{liu2022OSTC}.
    \item[4)] \textbf{Reference-model-assisted DRL without reducing action space:} We used single-period version of the reference-model-assisted SAC proposed in section \ref{subsection_WMA_SAC} except that the action space is $[\underline{\boldsymbol{a}}-\boldsymbol{a}_m, \bar{\boldsymbol{a}}-\boldsymbol{a}_m]$. The size of the action space was the same as the original DRL algorithm.
    \item[5)] \textbf{Reference-model-assisted DRL:} The simulation contained 11 experiments.
    We used the single-period version of the reference-model-assisted soft actor-critic algorithm proposed in section \ref{subsection_WMA_SAC}. We set $\boldsymbol{\delta}_o = (\bar{\boldsymbol{a}}-\underline{\boldsymbol{a}})/2$. For each experiment, $\boldsymbol{\delta} = \lambda_i * \boldsymbol{\delta}_o$, where $\lambda_i = 0,0.1,0.2,0.3,\dots 1$. 
    $\lambda_i = 0$ means the residual action space was $\O$, which was the same as the model-based optimization under a reference model.
\end{itemize}

The result of model-based optimization with an accurate power flow model was a baseline for evaluating the performance of DRL algorithms.
We trained the DRL agent using 300 days of data.
The hyper-parameters of three DRL algorithms were also the same as in our previous paper \cite{liu2022OSTC} as shown in Table \ref{DRL_paramter}. 
We tested the DRL algorithms in the same environment at each step in the training process. We stored the training results and testing results at each step.
In the simulation, we found the converged results of DRL for different random seeds are nearly equal, so we did not repeat each experiment for different random seeds.

\begin{figure*}[ht]
\centering
\subfigure[33-bus]{\begin{minipage}[t]{0.315\linewidth}\label{wm_DRL_result_33}
    \includegraphics[width=2.4in]{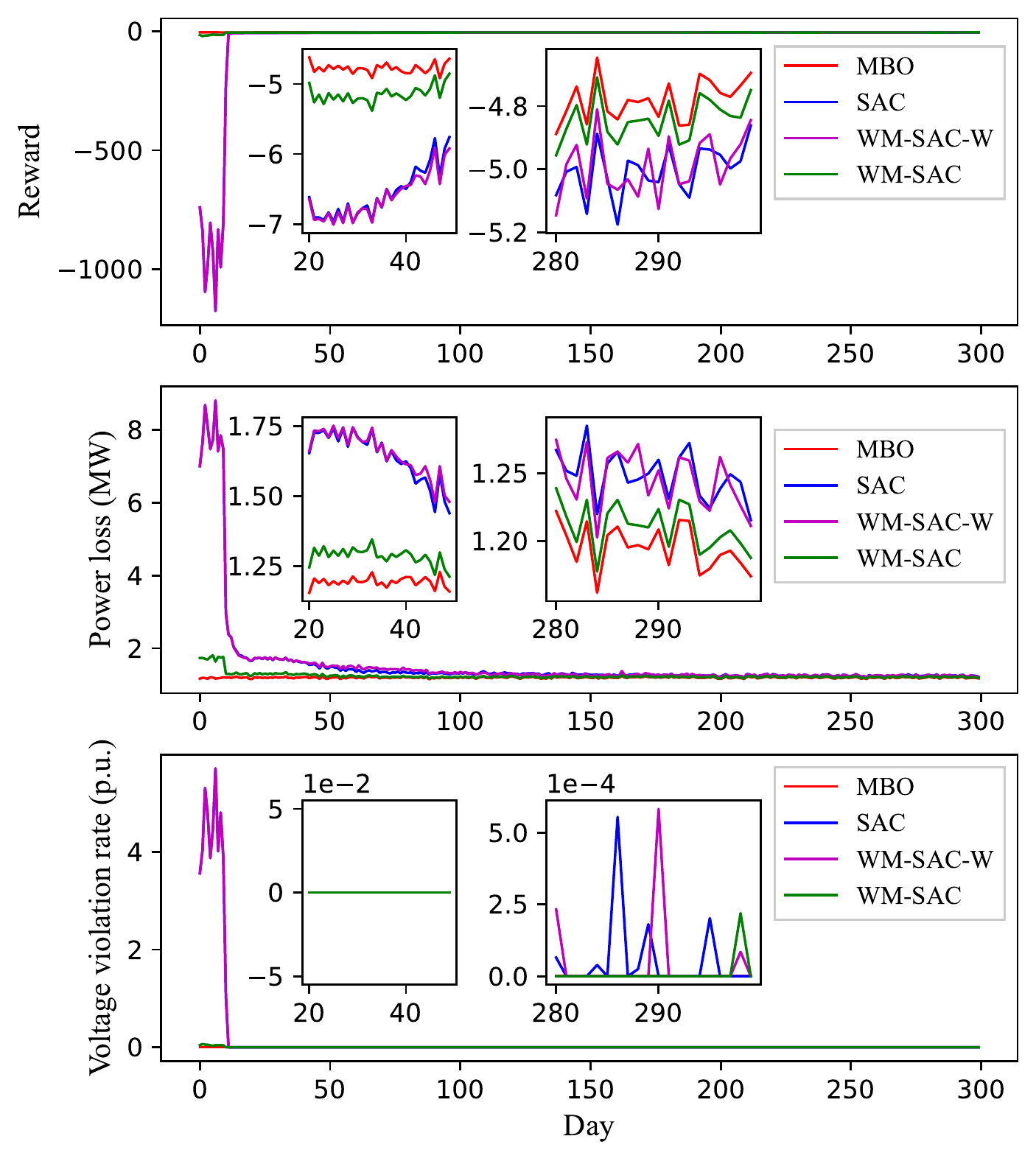}
    \end{minipage}
    }
\subfigure[69-bus]{
\begin{minipage}[t]{0.315\linewidth}\label{wm_DRL_result_69}
\includegraphics[width=2.4in]{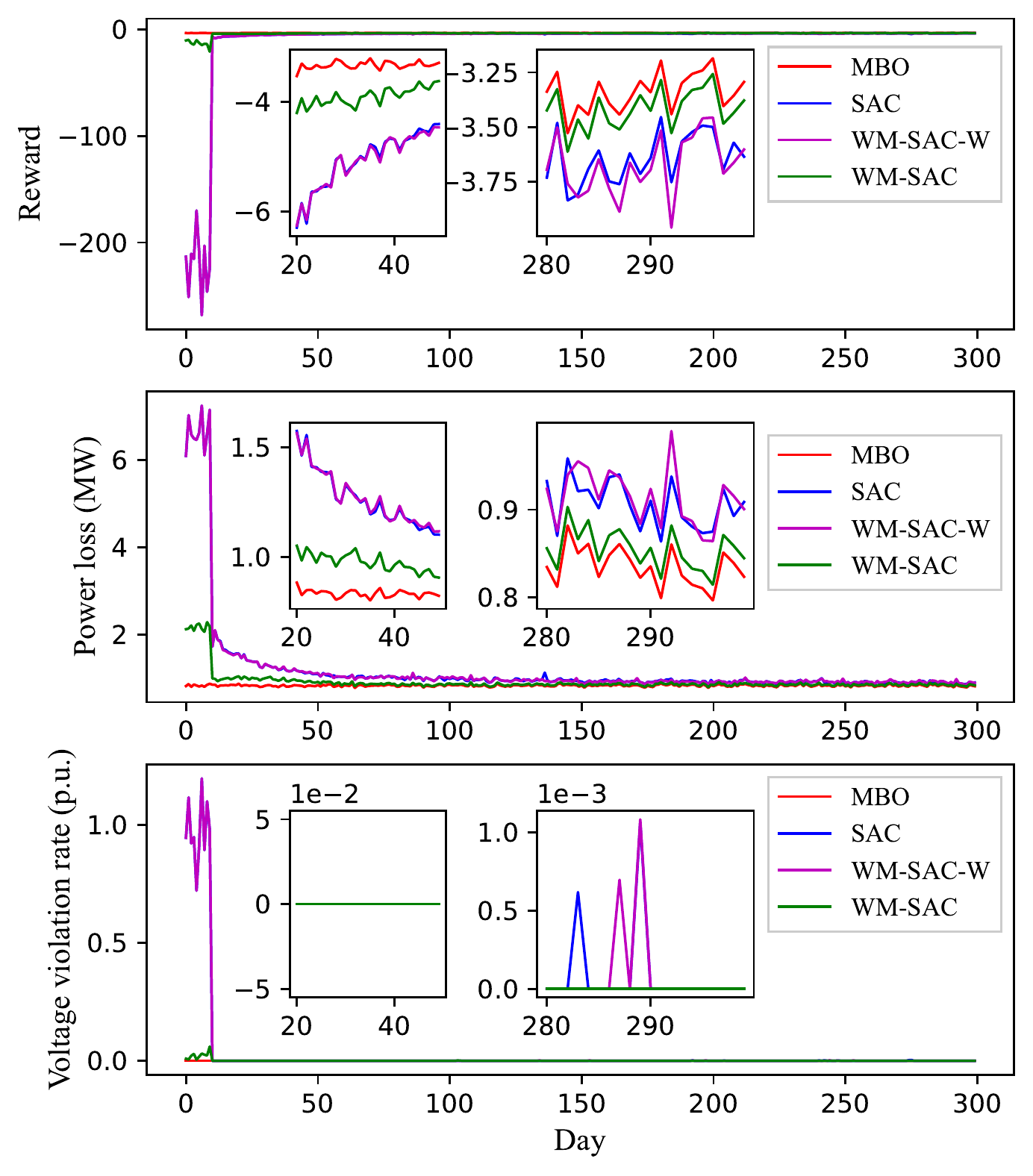}
\end{minipage}
    }
\subfigure[118-bus]{
\begin{minipage}[t]{0.315\linewidth}\label{wm_DRL_result_118}
\includegraphics[width=2.4in]{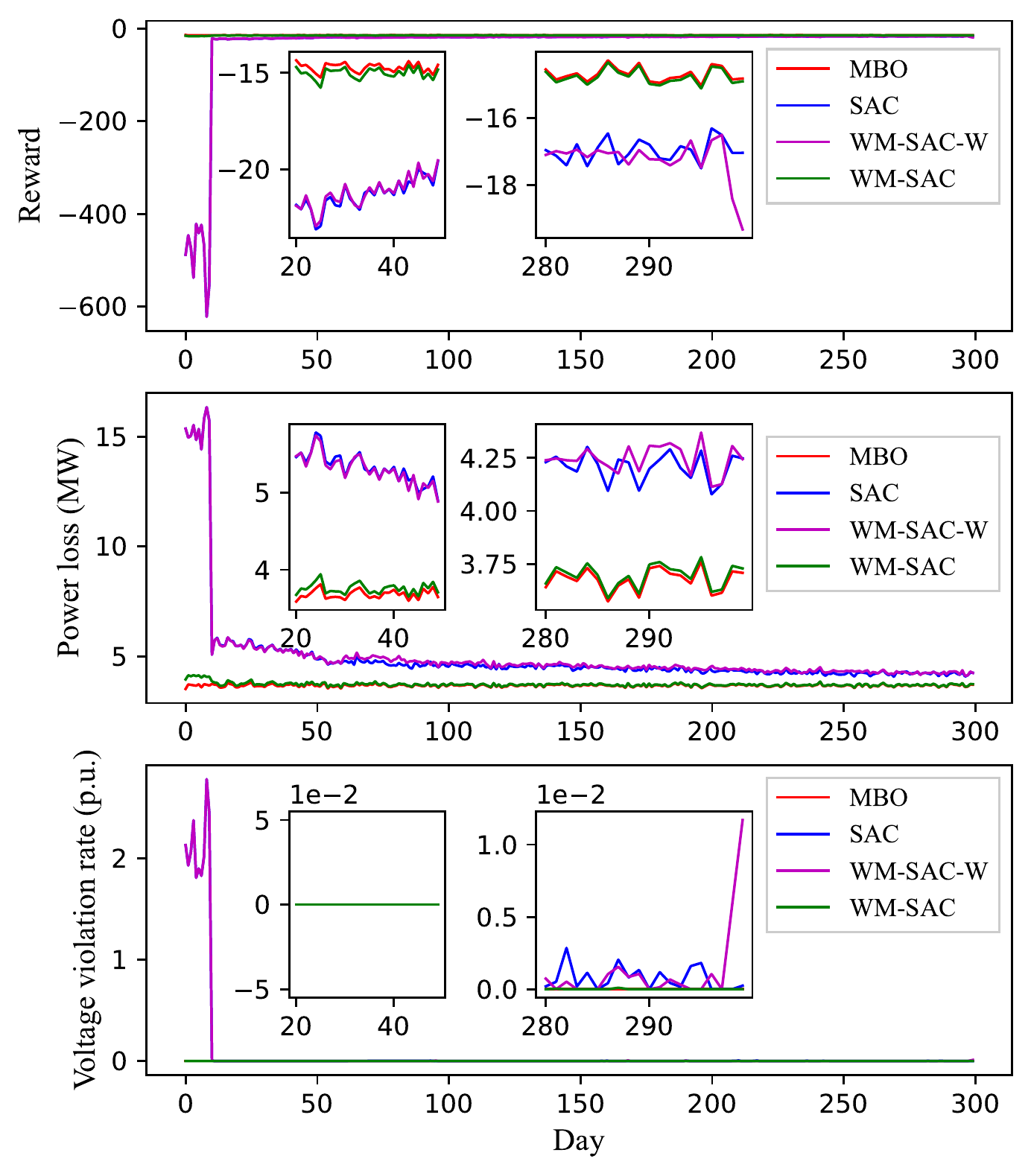}
\end{minipage}
    }
\caption{The testing results of the model-based optimization with an accurate model (MBO), SAC, reference-model-assisted SAC without reduced action space (WM-SAC-W) and reference-model-assisted SAC (WM-SAC) }\label{wm_DRL_result}
\end{figure*}

\begin{figure*}[ht!]
\centering
\subfigure[The reward error]{\begin{minipage}[t]{0.315\linewidth}\label{error_result_reward}
    \includegraphics[width=2.4in,height=1.1in]{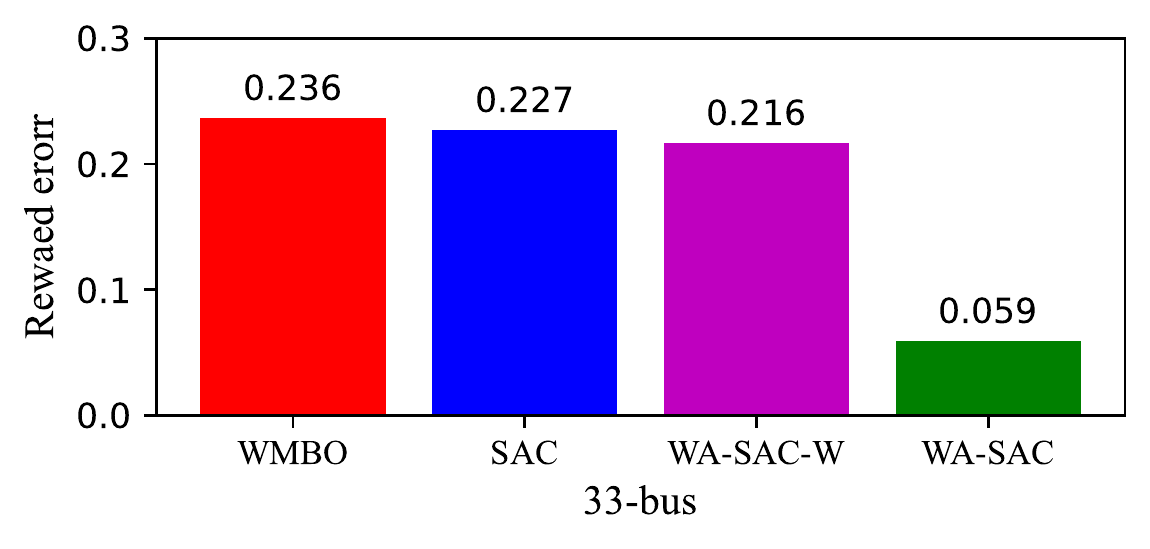}\\
    \includegraphics[width=2.4in,height=1.1in]{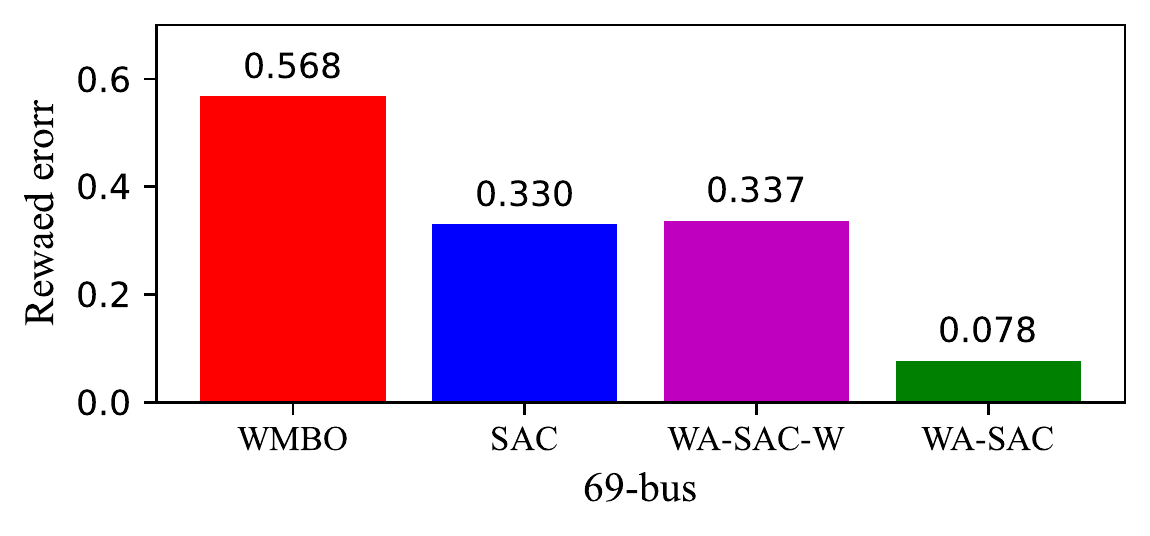}\\
    \includegraphics[width=2.4in,height=1.1in]{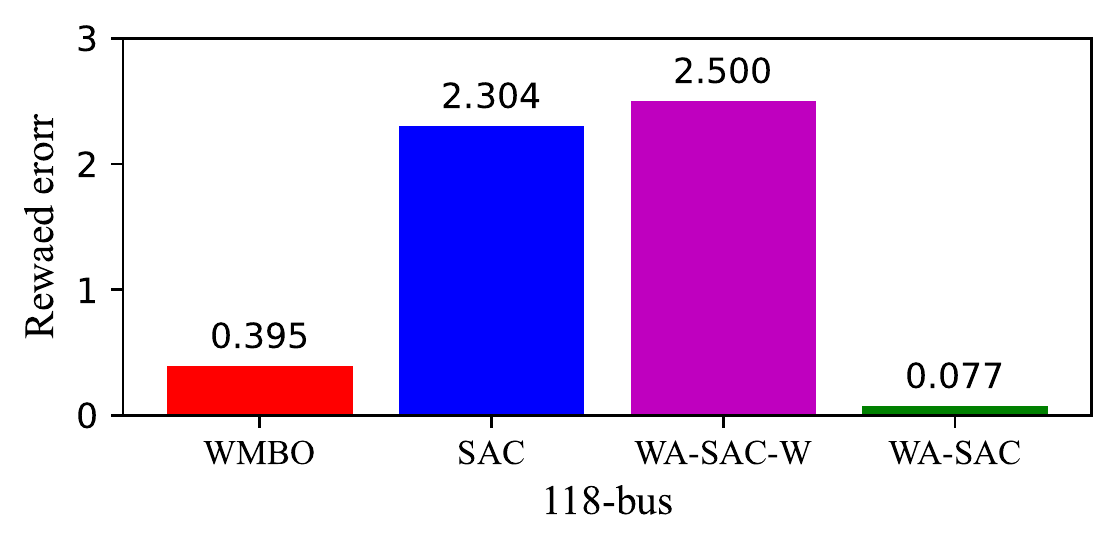}
    \end{minipage}
    }
\subfigure[The power loss error]{
\begin{minipage}[t]{0.315\linewidth}\label{error_result_power_loss}
\includegraphics[width=2.4in,height=1.1in]{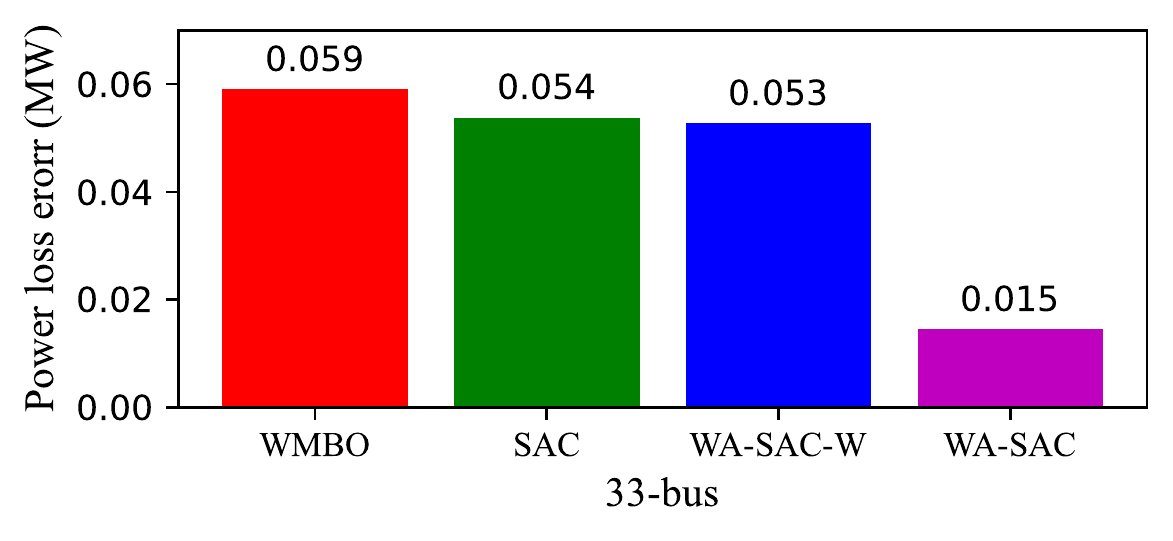}\\
\includegraphics[width=2.4in,height=1.1in]{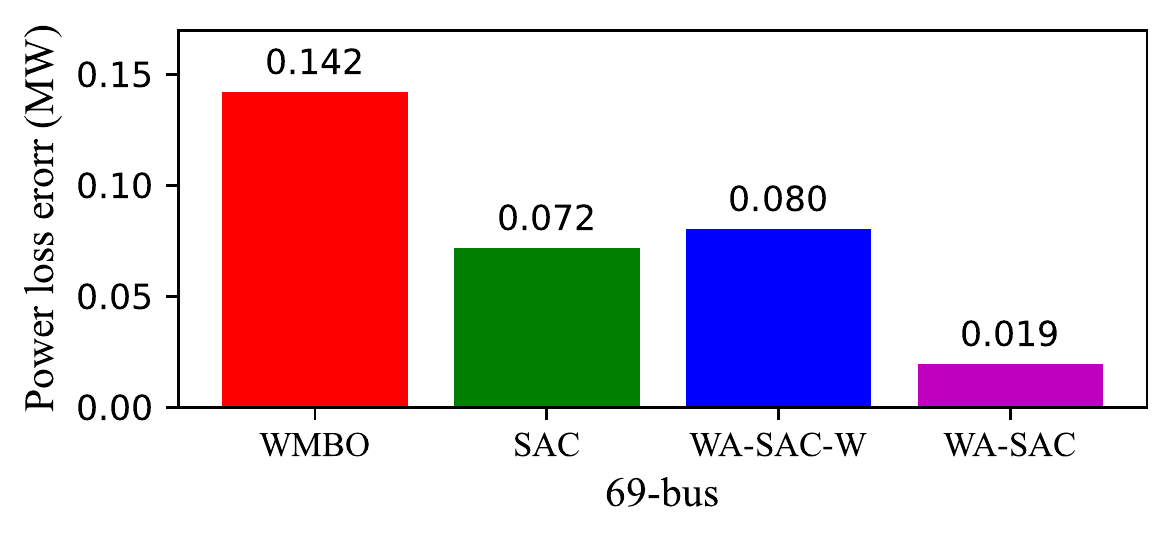}\\
\includegraphics[width=2.4in,height=1.1in]{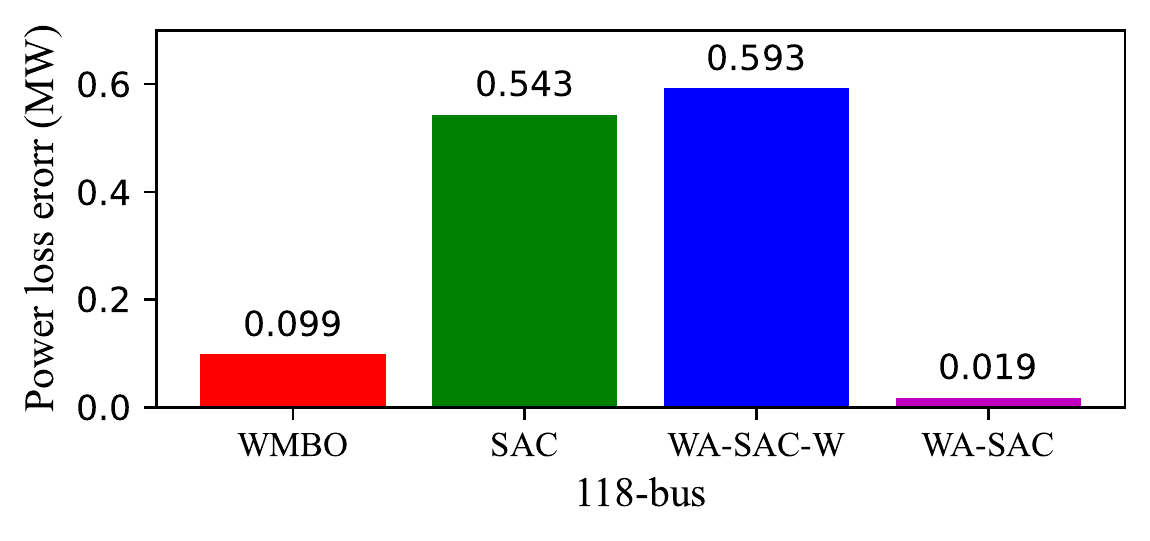}
\end{minipage}
    }
\subfigure[The voltage violation rate error ]{\begin{minipage}[t]{0.315\linewidth}\label{error_result_voltage}
    \includegraphics[width=2.4in,height=1.1in]{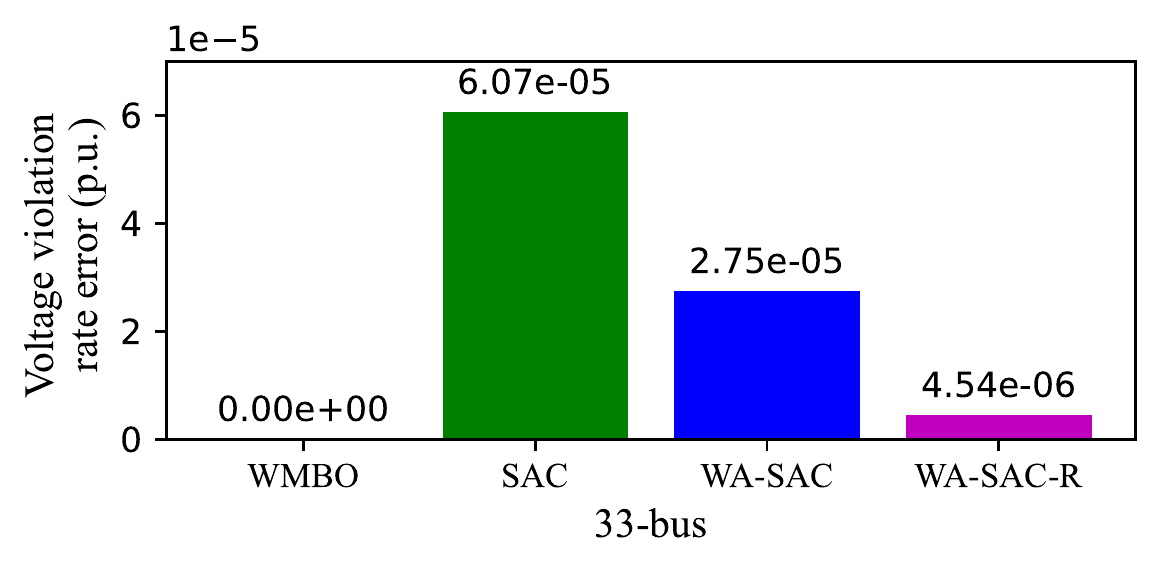}\\
    \includegraphics[width=2.4in,height=1.1in]{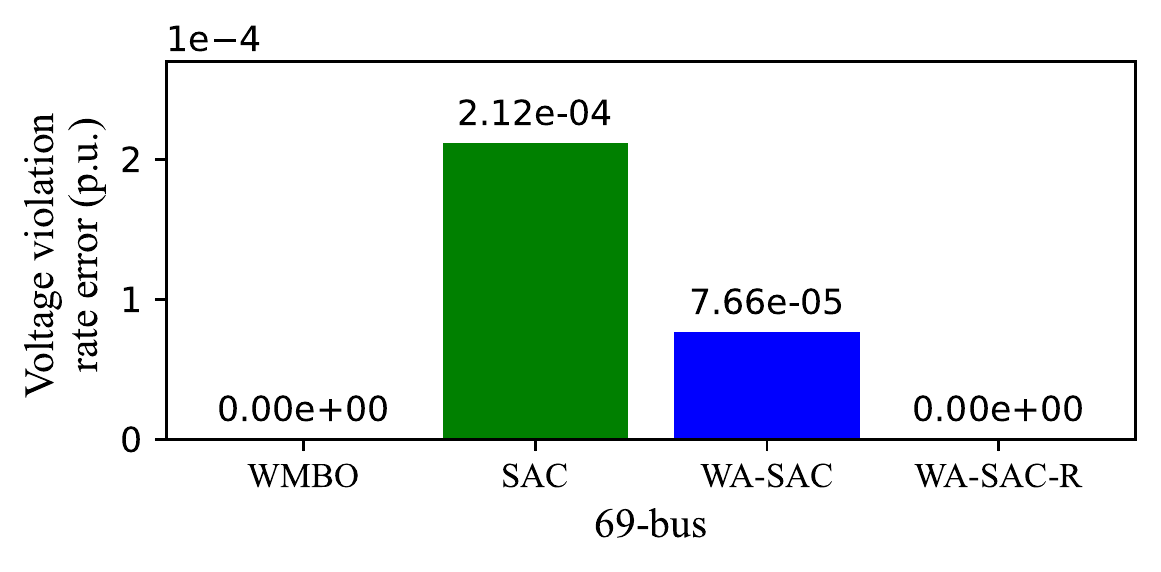}\\
    \includegraphics[width=2.4in,height=1.1in]{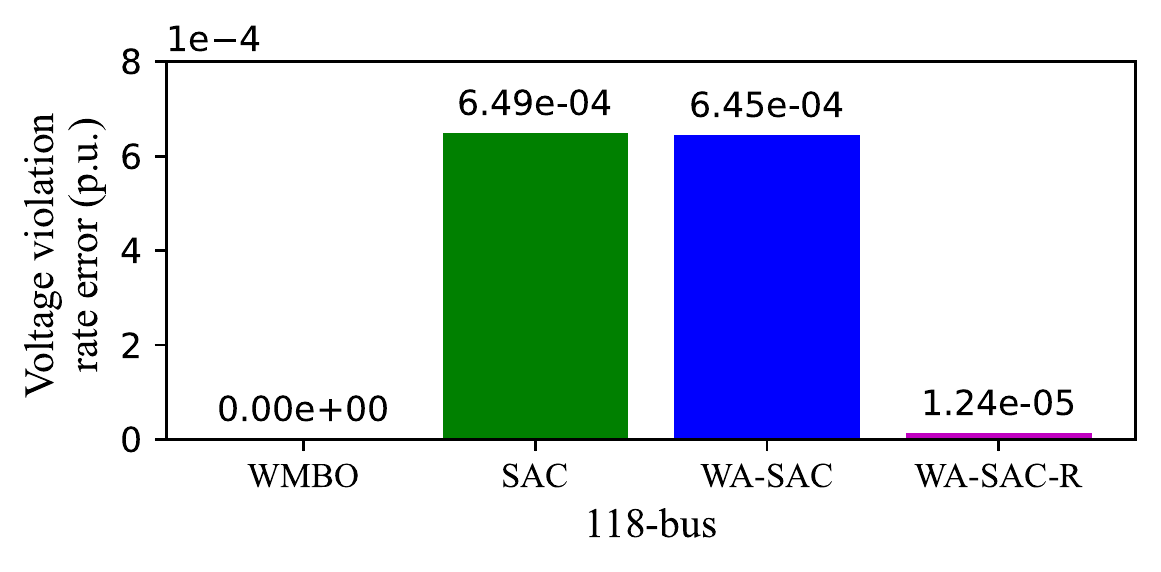}
    \end{minipage}
    }
\caption{The optimization result errors of the reference-model-based optimization (WMBO), SAC, reference-model-assisted SAC without reduced action space (WM-SAC-W) and reference-model-assisted SAC (WM-SAC) in the final 50 episodes. 
Here, the result error = the result of model-based optimization with an accurate model - the result of the mentioned method. }
\label{error_result}
\end{figure*}

\subsection{Verify Large Action Space Increases Learning Difficulties of DRL}

Five reasons for the ``too large" action space increasing learning difficulties of DRL proposed in section \ref{Proposition} and the effect of ``too small" action space of DRL discussed in section \ref{subsection_WMA_SAC} were verified by the results of experiment 5. The results are shown in Figs. \ref{wm1_large_sacle}, \ref{wm2_large_sacle}.
For clear presentation, we only plotted the results of $\lambda = 0.2,0.4,0.8 $ in Fig. \ref{wm1_large_sacle}. 
Fig.\ref{wm2_large_sacle} shows the 11 experiment results that the change of the critic loss, the training reward, the testing reward, and the training reward - the testing reward with the increasing of action space in the final 50 days.
The rewards in those figures were the daily accumulation values.

Five reasons for ``too large" action space increasing learning difficulties of DRL proposed in section \ref{Proposition} were verified by the following simulation phenomena:  
\begin{itemize}
    \item [1)] \textbf{For reason 1: Large action space exaggerates exploration noise:} Two experimental phenomena verified the reasons. Firstly, in the initial stage (the day 0-10 in Fig. \ref{wm1_large_sacle}), the action spaces increased, and the testing rewards decreased. 
    The actions of the day 0-10 were sampled from the exploration noise, so the corresponding rewards indicated the effect of the exploration noise.
    Secondly, in the final stage, with the increasing of  action space, the values of training reward minus testing reward of the final 50 days in Fig \ref{wm2_large_sacle} decreased. 
    The values shown the reward reduction due to exploration noise.
    
    \item [2)] \textbf{For reason 2: Large action space leads to the large critic error:}
    As shown the values of critic loss of the final 50 days in Fig \ref{wm2_large_sacle},
    the action space increased and the critic loss increased. The critic loss was the means of critic error for the sampling batch data in the final 50 days.
    
    \item [4)] \textbf{For reasons 3 and 4: The large critic error propagates to the actor} and \textbf{the large action space exaggerates the actor error} further:  In the final learning stage, the trajectory of testing reward increased and then decreased with increasing action space, as shown in Fig. \ref{wm2_large_sacle}. In the data of $\lambda_i =0.3,0.4,\dots, 1$ in Fig. \ref{wm2_large_sacle_33},  $\lambda_i = 0.5,0.6,\dots,1$ in Fig. \ref{wm2_large_sacle_69}, and $\lambda_i =0.2,0.3,\dots, 1$ in Fig. \ref{wm2_large_sacle_118}, the action space increased and the rewards decreased.
    These phenomenons indicates that the large critic error propagates to the actor and the large action space exaggerates the actor error further. 
    
    \item [5)] \textbf{For reason 5: The coupling of actor and critic slows down the convergence of DRL:}
    As shown in the testing reward trajectory of the day $20-50$ in Fig. \ref{wm1_large_sacle}, the action space increased, and the DRL algorithms converged more slowly.
\end{itemize}

The effect of the ``too small" action space of DRL discussed in section \ref{subsection_WMA_SAC} was verified also in Fig. \ref{wm2_large_sacle}.
In the data of $\lambda_i = 0,0.1,0.2, 0.3$ for 33-bus distribution network, $\lambda_i = 0,0.1,\dots 0.5$ for 69-bus distribution network, and $\lambda_i = 0,0.1,0.2$ for 118-bus distribution network, testing rewards increased with the increasing of action space. It indicates the problem of ``too small"  action space that the final actions cannot reach the optimal actions. Increasing the action space alleviated the problem and made the actions more likely to close optimal actions.


\subsection{Verify the Superiority of Reference-Model-Assisted DRL}

To verify the superiority of reference-model-assisted DRL,
we plotted the results of DRL, reference-model-assisted DRL without reducing action space, and reference-model-assisted DRL in Fig. \ref{wm_DRL_result}. 
We selected $\lambda = 0.3$ in the 33-bus distribution network, $\lambda = 0.5$ in the 69-bus distribution network, and $\lambda = 0.2$ in the 118-bus distribution network, because their performance was the best among the experiments of reference-model-assisted DRL.

We made two observations. 

First and foremost, reference-model-assisted SAC had a better performance than SAC and reference-model-assisted SAC without reducing action space in terms of reward, power loss, and violation rate in the training process. In contrast, reference-model-assisted SAC without reducing action space had a similar performance as DRL. It verifies that the effectiveness of reference-model-assisted DRL is only because it reduces the action space.

Second, in the random exploration stage and the initial learning stage, the performance of reference-model-assisted SAC was considerably better than SAC from the view of reward, power loss, and violation rate, whereas just slightly worse than the model-based optimization with an accurate model. This indicates that the reference-model-assisted DRL is more likely to be trained in real ADNs because of the safety issues.

Fig. \ref{error_result} shows the quantified optimization results error of the four methods compared with the baseline results of the model-based optimization method with an accurate model. From the perspective of reward and power loss, reference-model-assisted SAC improved control accuracy at least 3.6 times on the 33-bus distribution network and the 69-bus distribution network, and 29 times on the 118-bus distribution network compared with SAC and reference-model-assisted SAC without reducing action space.
From the perspective of voltage violation rate, reference-model-assisted SAC improved control accuracy about 6 times on the 33-bus distribution network and 52 times on the 69-bus distribution network compared with SAC and reference-model-assisted SAC without reducing action space.
Even reference-model-assisted SAC cannot eliminate the voltage violation theoretically, it achieved no voltage violation in the experiment.

\section{Conclusion}
We first analyzed that large action space increased the learning difficulties of DRL algorithms and then proposed a reference-model-assisted DRL approach to reduce the action space. In particular, five essential reasons were suitably provided from the process of generating data and training neural networks to explain the demerits of large action space. Those reasons were demonstrated point-by-point in the simulations. We analyzed two potential problems of roughly setting residual action space and gave the intuition explanation and solution. Corresponding simulations show the superiority of the proposed method, and the main reason for the superiority is it reduces the action space.

The proposed method is a general and comprehensive method for constrained optimization problems. In the future, we will extend the method to more optimization problems in the power system field, to achieve more desirable outcomes in real-world engineering conditions.

\ifCLASSOPTIONcaptionsoff
  \newpage
\fi



\bibliographystyle{IEEEtran}
\bibliography{IEEEabrv,ref.bib}
\end{document}